\documentclass[twocolumn]{autart} 
\usepackage{amssymb}
\usepackage{amsmath}
\usepackage{color}
\usepackage{graphics}
\usepackage{graphicx}
\usepackage{multirow}
\usepackage{algorithm,algorithmic}
\usepackage{savesym}
\savesymbol{AND}
\usepackage{epstopdf}
\usepackage{subcaption}
\usepackage{tikz}
\usetikzlibrary{arrows}
\usepackage{natbib}
\usepackage{url}

\def \R{\mathbb{R}}

\def \Z{\mathbb{Z}}

\newtheorem{theorem}{Theorem}
\newtheorem{remark}{Remark}
\newtheorem{lemma}{Lemma}
\newtheorem{definition}{Definition}

\newtheorem{proposition}{Proposition}

\newtheorem{assumption}{Assumption}
\newtheorem{example}{Example}

\begin{document}

\begin{frontmatter}

\title{Computation of the maximal invariant set of discrete-time linear systems subject to a class of non-convex constraints}

\author[AddA]{Zheming Wang} \ead{zheming.wang@uclouvain.be},
\author[AddA]{Rapha\"el M. Jungers} \ead{raphael.jungers@uclouvain.be},
\author[AddB]{Chong Jin Ong}\ead{mpeongcj@nus.edu.sg}

\thanks{Rapha\"el M. Jungers is a FNRS Research Associate. He is supported by the French Community of Belgium, the Walloon Region and the Innoviris Foundation.}
\thanks{A preliminary version was presented at the 17th European Control Conference (ECC), 2019.}

\address[AddA]{The ICTEAM Institute, UCLouvain, Louvain-la-Neuve,1348, Belgium}
\address[AddB]{Department of Mechanical Engineering, National University of Singapore, 117576, Singapore}


\begin{abstract}
We consider the problem of computing the maximal invariant set of discrete-time linear systems subject to a class of non-convex constraints that admit quadratic relaxations. These non-convex constraints include semialgebraic sets and other smooth constraints with Lipschitz gradient. With these quadratic relaxations, a sufficient condition for set invariance is derived and it can be formulated as a set of linear matrix inequalities. Based on the sufficient condition, a new algorithm is presented with finite-time convergence to the actual maximal invariant set under mild assumptions.  This algorithm can be also extended to switched linear systems and some special nonlinear systems. The performance of this algorithm is demonstrated on several numerical examples.
\end{abstract}

\end{frontmatter}

\section{Introduction}

Invariant set theory is an important tool for stability analysis and controller design of constrained dynamical systems. This theory has been used to solve various problems in systems and control; see, for instance,  \citep{BOO:A91,ART:Bla99,BOO:BM08,BOO:BYG17} and the references therein. An invariant set of a dynamical system refers to a region where the trajectory will never leave once it enters. One well-known application is in Model Predictive Control (MPC) \citep{ART:MRRS00}, where invariant sets are often used to ensure recursive feasibility and stability.


Given the extensive applications of invariant sets in systems and control,  significant attention has been paid to their characterization and computation. In \citep{ART:GT91,ART:DH99,INP:PRSD05}, recursive algorithms have been proposed to compute polyhedral invariant sets of linear systems. For linear systems with bounded disturbances, robust invariant sets can be computed using different algorithms \citep{ART:KG98,ART:RKKM05,ART:OG06,ART:RKMK07,ART:AR08,ART:P16}. For linear systems with control, the computation of (control) invariant sets is more complicated and a few algorithms have been proposed to compute inner or outer approximations \citep{ART:GC87,ART:DC17,ART:RT17}. Algorithms for computing invariant sets of different nonlinear systems are also available in the literature, see, e.g., \citep{ART:BLAC05,ART:ACFC09,ART:FAC10,ART:SG12,ART:HK14,ART:KHJ14}. The concept of set invariance can be extended to hybrid systems.  For instance, invariant sets can be defined for switched systems, which constitute an important family of hybrid systems, and the computation of such sets have been extensively studied, see, e.g., \citep{ART:DO12,ART:HTY16,INP:AJ16,ART:ASJ17,ART:AR18,ART:LTJ18}.

Among various invariant sets, the maximal invariant set is of particular interest. A standard algorithm for computing the maximal invariant set of linear systems with polytopic constraints is presented in \citep{ART:GT91,ART:KG98} with sufficient conditions for finite convergence. Since recently, necessary and sufficient conditions for finite convergence have been well understood \citep{ART:AG18}. Even though the literature on set invariance of linear systems is large, computing the exact maximal control invariant set is still challenging, especially when the constraints are non-convex, see, e.g., recent works \citep{ART:DC17,ART:RT17} for inner or outer approximations. For switched linear systems, algorithms to compute the maximal invariant set are also provided in the cases of polytopic/convex constraints \citep{ART:DO12,ART:DO12b,ART:ASJ17,ART:AR18} and semialgebraic constraints \citep{INP:AJ16}. Although there are some algorithms for estimating the maximal invariant sets of certain types of nonlinear systems, see, e.g., \citep{ART:ACFC09,ART:HK14,ART:KHJ14}, computing the exact maximal invariant set is still an open problem for general nonlinear systems. When the constraints are non-convex, the computation will be even more challenging. In fact, in the presence of non-convex constraints, to the best of our knowledge, the exact computation of the maximal invariant set is only addressed in \citep{INP:AJ16} for switched linear systems with semialgebraic constraints by lifting the original system into a higher dimension. For general non-convex constraints, computing the exact maximal invariant set is an unsolved problem even for linear systems.

This paper is focused on the exact computation of the maximal invariant set of discrete-time linear systems in the presence of a broad class of non-convex constraints that admit quadratic relaxations. We subsequently generalize our method to some classes of nonlinear systems. We will give formal assumptions on such non-convex constraints which include semialgebraic constraints and smooth constraints with Lipschitz gradient. Using quadratic relaxations, a sufficient condition for set invariance is derived from the S-procedure \citep{BOO:BGFB94} and can be expressed as a set of Linear Matrix Inequalities (LMI). Based on this sufficient condition, we present a new algorithm that solves a set of LMIs at each iteration. The tightness of the sufficient condition largely depends on the conservatism of the S-procedure \citep{ART:DP06}. We emphasize that, even though the S-procedure induces some conservatism in the sufficient condition, our algorithm converges to the \emph{true} maximal invariant set in finite time, as we show below. Moreover, as we show on several examples, the algorithmic efficiency of our technique turns out to be much better than the previously known techniques in the literature. This proposed algorithm can be also extended to switched linear systems and some nonlinear systems that can be linearized via state transformation. In the case of semialgebraic constraints, a similar lifting method as \citep{INP:AJ16} is used. The dimension of the lifted space depends on the order of the semialgebraic constraints. It will be shown that we require a lower lifted system than \citep{INP:AJ16} for the same setting.

A preliminary version of this paper appears as a conference paper in \citep{INP:WJO19}, which is only focused on linear systems. In this paper, we provide complete detailed proofs of all lemmas and theorems, the discussion on the extensions to switched linear systems and some special nonlinear systems, and additional numerical results.

The rest of the paper is organized as follows. This section ends with the notation,  followed by the next section on the review of preliminary results on the invariant sets of linear systems. Section \ref{sec:main} presents the proposed approach for computing the maximal invariant set of linear systems with non-convex constraints. Section \ref{sec:nlsys} discusses semi-algebraic constraints and the extensions some special nonlinear systems. Several numerical examples are provided in Section \ref{sec:num}. The last section concludes the work.

The notation used in this paper is as follows. Non-negative and positive integer sets are indicated respectively by $\mathbb{Z}^+_0$ and $\mathbb{Z}^+$. Similarly, $\mathbb{R}^+_0$ and $\mathbb{R}^+$ refer respectively to the sets of non-negative and positive real numbers.  For any $M\in \mathbb{Z}^+$, let $\mathcal{I}_M:=\{1,2,\cdots,M\}$. For any given set $S=\{s_1,s_2,\cdots,s_M\}$, $cone(S)$ denotes the positive linear span of $S$, i.e., $cone(S):=\{\sum_{i=1}^M \alpha_i s_i: \alpha_i \in  \mathbb{R}^+_0, i\in \mathcal{I}_M\}$.  $\mathbb{S}^n$ denotes the set of symmetric matrices in $\mathbb{R}^{n\times n}$. $I_n$ (the subscription is omitted when the dimension is clear from the context) is the $n\times n$ identity matrix and $\pmb{1}_n$ denote the vector of $n$ ones.  For a square matrix $Q$, $Q \succ (\succeq) ~ 0$ means $Q$ is positive definite (semi-definite). The $p$-norm of $x\in \mathbb{R}^{n}$ is $\|x\|_p$ while $\|x\|^2_Q=x^TQx$ for $Q \succeq 0$. Given a set of vectors, $x_i \in \mathbb{R}^{n_i}, i \in \mathcal{I}_M$, the collection of vectors, $(x_1, x_2, \cdots, x_M)$ also refers to the stack vector of $[(x_1)^T ~ (x_2)^T ~ \cdots ~ (x_M)^T]^T \in \mathbb{R}^{\sum_{i=1}^M n_i}$ for notational simplicity. Additional notation is introduced as required in the text.

\section{Preliminaries}\label{sec:pre}
This section reviews some known results on the invariant sets of constrained discrete-time linear systems. We consider the linear system
\begin{align}
x(t+1) = Ax(t), \quad \forall t \in \mathbb{Z}^+_0, \label{eqn:xA}
\end{align}
where $x(t) \in \mathbb{R}^n$ is the state vector. The system is subject to state constraints
\begin{align}\label{eqn:xXt}
x(t) \in X:=\Omega \bigcap \Theta, \quad  \forall t \in \mathbb{Z}^+_0.
\end{align}
where $\Omega \subseteq \mathbb{R}^n$ is a quadratic set and $\Theta \subseteq \mathbb{R}^n$ is a set of non-quadratic nonlinear constraints. The set $\Omega$ is described as
\begin{align}\label{eqn:omegaQi}
\Omega = \{x\in \mathbb{R}^n : x^TQ_i x+2q_i^Tx\le 1, i \in \mathcal{I}_p\}, 
\end{align}
where $Q_i\in \mathbb{S}^n$, $q_i\in \mathbb{R}^n$ and $p$ is the number of constraints. When $Q_i=0$, for all $i \in \mathcal{I}_p$, $\Omega$ becomes a polytope. The set $\Theta$ is described as
\begin{align}\label{eqn:xtheta}
\Theta :=\{x\in \mathbb{R}^n: H_{i}(x) \le 1, i \in \mathcal{I}_m\}
\end{align}
where $H_{i}: \mathbb{R}^n \rightarrow \mathbb{R}$ is a continuous nonlinear function and $m \in \mathbb{Z}^+$ is the number of such nonlinear constraints. 

For computational reasons, we treat quadratic constraints and general nonlinear constraints differently. The following assumptions are made.

\begin{assumption}\label{ass:A}
The matrix $A$ is Schur stable, i.e., for any eigenvalue $\lambda$ of A, $|\lambda|$ is smaller than one.
\end{assumption}

\begin{assumption}\label{ass:interior}
The set $\Omega$ is compact and contains the origin in its interior. 
\end{assumption}

\begin{assumption}\label{ass:smooth}
For any $i \in \mathcal{I}_m$,  $H_{i}: \mathbb{R}^n \rightarrow \mathbb{R}$ is a continuous nonlinear function with $H_i(0)=0$ and there exist a vector $H_{i}^{\nabla}\in \mathbb{R}^n$ and a scalar $L_i \ge 0$ such that
\begin{align}\label{eqn:Hi0x}
|H_i(x)-H_i(0) - (H_{i}^{\nabla})^T x|\le \frac{L_i}{2} \|x\|^2
\end{align}
for all $x\in \Omega$.
\end{assumption}


Assumptions \ref{ass:A} and \ref{ass:interior} are standard requirements that are often made in the literature, see, e.g., \citep{ART:GT91}. From the continuity of the nonlinear functions $\{H_i(x)\}_{i=1}^m$, $\Theta$ contains the origin in its interior, and thus $X$ is compact and contains the origin in its interior. Assumption \ref{ass:smooth} requires all the nonlinear functions to have quadratic lower and upper bounds. However, these functions are not necessarily Lipschitz continuous or differentiable. Clearly, for functions with Lipschitz continuous gradient, the condition in Assumption \ref{ass:smooth} will be satisfied. Indeed, suppose that, for any $i \in \mathcal{I}_m$, $H_{i}$ is a continuously differentiable function with Lipschitz gradient:
\begin{align}
\|\nabla H_{i}(x)-\nabla H_{i}(y)\| \le L_{i}\|x-y\|, \forall x,y \in \Omega,
\end{align}
then, Assumption \ref{ass:smooth} is satisfied with $H_{i}^{\nabla}=\nabla H_i(0)$ (see, e.g., Lemma 6.9.1 in \citep{BOO:B09}).  Inspired by a recent work on different classes of quadratic approximations \citep{ART:NNG19},   we will refer to a function satisfying (\ref{eqn:Hi0x}) as a \emph{quasi-smooth} function. All the polynomial functions satisfy (\ref{eqn:Hi0x}). For notational simplicity, a compact form of $\Theta$ is given below
\begin{align}
\Theta = \{x\in \mathbb{R}^n: H(x) \le \pmb{1}_m\}
\end{align}
where $
H(x):=(H_1(x),H_2(x),\cdots, H_m(x)).
$


We now define some central concepts of this paper.

\begin{definition}
\citep{ART:Bla99,ART:MRRS00} The nonempty set $Z\subseteq X$ is a \emph{CA-invariant} (Constraint Admissible invariant) set for System (\ref{eqn:xA})  if for any $x\in Z$ one has that $Ax\in Z$. 
\end{definition}

With Assumptions \ref{ass:A} and \ref{ass:interior}, there often exist multiple \emph{CA-invariant} sets. In many applications, it is desirable to compute the maximal \emph{CA-invariant} set \citep{ART:GT91}, which is defined below.

\begin{definition}
A nonempty set $S\subseteq X$ is the maximal \emph{CA-invariant} set for the system (\ref{eqn:xA}) if $S$ is a \emph{CA-invariant} set and contains all \emph{CA-invariant} sets in $X$.
\end{definition}
It is a standard result that the maximal \emph{CA-invariant} set exists (see \citep{ART:GT91} for general conditions guaranteeing its existence), and that it can be computed recursively by the following iteration:
\begin{align}
O_0 &:= X, \label{eqn:O0}\\
O_{k+1} &:= O_k \bigcap \{x\in \mathbb{R}^n: Ax\in O_k\}, k \in \mathbb{Z}^+_0 . \label{eqn:Ok}
\end{align}
With these iterates, it can be verified that
\begin{align}\label{eqn:OkA}
O_k = \{x\in X: A^\ell x \in X, \ell \in \mathcal{I}_k\}, k \in \mathbb{Z}^+.
\end{align}
Thus, the maximal \emph{CA-invariant} set can be expressed as
\begin{align}\label{eqn:Oinf}
O_{\infty}:=\bigcap_{k \in \mathbb{Z}^+_0} O_{k} = \{x\in \mathbb{R}^n: A^k x\in X, k \in \mathbb{Z}^+_0\}.
\end{align}
From Assumptions \ref{ass:A} and \ref{ass:interior}, the set $O_{\infty}$ defined in (\ref{eqn:Oinf}) has the following properties \citep{ART:GT91}: (i) if $Z\subseteq \mathbb{R}^n$ is a \emph{CA-invariant} set of system (\ref{eqn:xA}), $Z\subseteq O_{\infty}$; (ii) there exists a finite $k^*$ such that $
O_{k^*+1}=O_{k^*}
$; (iii) for any $k^*$ satisfying (ii), it can be shown that
$
O_{k}=O_{k^*}
$
for all $k \ge k^*$ and $O_{\infty} = O_{k^*}$.

From the properties above, the problem of computing $O_{\infty}$ becomes the search for an index $k^*$ such that $O_{k^*+1}=O_{k^*}$. The standard procedure is to increase $k$ from $0$ until
$
O_{k+1} =  O_{k},
$
which is equivalent to
\begin{align}\label{eqn:OkAX}
O_{k} \subseteq \{x\in \mathbb{R}^n: A^{k+1}x\in X\},
\end{align}
see \citep{ART:GT91} for details. This condition can be treated as a stopping criterion for the algorithm in (\ref{eqn:O0})-(\ref{eqn:Ok}). Observe that $\{x\in \mathbb{R}^n: A^{k+1}x\in X\}$ can be rewritten as
$
\{x\in \mathbb{R}^n: (A^{k+1}x)^TQ_iA^{k+1}x +2q^T_iA^{k+1}x\le 1,
i\in \mathcal{I}_p, H(A^{k+1}x) \le \pmb{1}_m \}, \forall k \in \mathbb{Z}^+_0.
$
During the computational procedure, we aim to find the minimal $k$ that satisfies (\ref{eqn:OkAX}). Let 
\begin{align}
k_{\min}:=  \arg \min_{k\in \mathbb{Z}^+_0} \{k: (\ref{eqn:OkAX}) \textrm{ holds} \}.
\end{align}
As shown in Property (iii), $O_{k}=O_{k_{\min}}=O_{\infty}$ for any $k\ge k_{\min}$. By this property,  given any upper bound on $k_{\min}$, one is able to determine $O_{\infty}$. When there are only linear constraints, the standard algorithm for the verification of (\ref{eqn:OkAX}) is to solve a set of linear optimization problems, see, e.g., \citep{ART:Bla99}. However, in the presence of non-convex constraints, we need to solve a set of nonlinear optimization problems, which are computationally expensive. For this reason, we will aim to derive a sufficient condition that can be efficiently verified.

\section{The proposed approach}\label{sec:main}
This section discusses the computation of the exact maximal \emph{CA-invariant} set with nonlinear constraints.  An algorithm will be presented to compute an upper bound on $k_{\min}$ which can be determined in a finite number of iterations under mild assumptions.

For quadratic (or linear) constraints, the following nonlinear optimization problem is defined at the $k^{th}$ iteration of (\ref{eqn:Ok}):
\begin{subequations}\label{eqn:gmax}
\begin{align}
g_i(k) : = &\max\limits_{x} (A^{k+1}x)^TQ_i A^{k+1}x+2q_i^TA^{k+1}x \\
\textrm{s.t.} \quad & x \in O_k
\end{align}
\end{subequations}
for $i \in \mathcal{I}_p$ and let $g_{\max}(k) :=\max_{i\in \mathcal{I}_p} g_i(k)$. If $g_{\max}(k) \le 1$ for some $k \in \mathbb{Z}^+_0$, $O_k \subseteq \{x\in \mathbb{R}^n:  (A^{k+1}x)^TQ_iA^{k+1}x+2q^T_ix\le 1, i\in \mathcal{I}_p\}$. Similarly, for non-quadratic nonlinear constraints, the following nonlinear optimization problem is defined at the $k^{th}$ iteration of (\ref{eqn:Ok}):
\begin{subequations}\label{eqn:hmax}
\begin{align}
h_i(k) : = &\max\limits_{x} H_i(A^{k+1}x)\\
\textrm{s.t.} \quad & x \in O_k
\end{align}
\end{subequations}
for $i\in \mathcal{I}_m$ and $h_{\max}(k) :=\max_{i\in \mathcal{I}_m} h_i(k)$. If $h_{\max}(k)\le 1$ for some $k \in \mathbb{Z}^+_0$, $O_k \subseteq\{x\in \mathbb{R}^n: H(A^{k+1}x) \le \pmb{1}_m\} $. Using (\ref{eqn:gmax}) and (\ref{eqn:hmax}), $k_{\min}$ can be determined via $\min_{k\in \mathbb{Z}^+_0} \{k:g_{\max}(k)\le 1, h_{\max}(k)\le 1\}$. To do so, we need in principle to solve (\ref{eqn:gmax}) and (\ref{eqn:hmax}) and get their global optimal solutions. However, for general nonlinear constraints, both (\ref{eqn:gmax}) and (\ref{eqn:hmax}) are nonlinear non-convex problems.  Even if $\Omega$ and $\Theta$ are convex sets, (\ref{eqn:gmax}) and (\ref{eqn:hmax}) may not be convex problems. Therefore, we only require upper bounds on the optimal values of $g_{\max}(k) $ and $h_{\max}(k) $.

\subsection{Quadratic constraints}\label{sec:quad}
Consider the case where only quadratic constraints exist, i.e., $\Theta=\mathbb{R}^n$ and $X=\Omega$. Let
\begin{align}
\bar{A} &= \left(\begin{array}{cc} A & 0\\ 0 & 1 \end{array}\right) \textrm{ and } \label{eqn:barA}\\
\bar{Q}_i :& = \left(\begin{array}{cc} Q_i & q_i\\ q_i^T& -1 \end{array}\right), \forall i \in \mathcal{I}_p. \label{eqn:barQi}
\end{align}
Following the iteration in (\ref{eqn:O0})-(\ref{eqn:Ok}), we define:
\begin{align}
\mathcal{Q}_0 &:= \{\bar{Q}_i, i\in \mathcal{I}_p\} \label{eqn:mathcalQ0}\\
\mathcal{Q}_{k+1} &:= \mathcal{Q}_k \bigcup \bar{A} ^T\mathcal{Q}_k \bar{A}, k\in \mathbb{Z}^+_0 \label{eqn:mathcalQk}
\end{align}
where $\bar{A} ^T\mathcal{Q}_k \bar{A}:= \{\bar{A} ^T\bar{Q} \bar{A}: \bar{Q}\in \mathcal{Q}_k\}$. From the construction of $\mathcal{Q}_{k}$, it can be shown that
\begin{align}
\mathcal{Q}_{k} & \subseteq  \{\bar{Q}_1,\cdots, \bar{Q}_p, \bar{A}^T\bar{Q}_1\bar{A},  \cdots, \bar{A}^T\bar{Q}_p\bar{A},\cdots,\nonumber\\
  &(\bar{A}^{k})^T\bar{Q}_1 \bar{A}^{k}, \cdots, (\bar{A}^{k})^T\bar{Q}_p \bar{A}^{k} \}, \quad k\in \mathbb{Z}^+_0,
\end{align}
with $|\mathcal{Q}_{k}|  \le  (k+1)p$. It can be also shown that 
\begin{align}
\mathcal{Q}_{k+1}\setminus \mathcal{Q}_{k}  \subseteq  \{(\bar{A}^{k+1})^T\bar{Q}_i \bar{A}^{k+1}, i \in \mathcal{I}_p\} \label{eqn:deltaQk}
\end{align}
for all $k\in \mathbb{Z}^+_0$. Using the notation above, $O_k$ defined in  (\ref{eqn:O0})-(\ref{eqn:Ok}) can be rewritten as
\begin{align}\label{eqn:OkmathcalQk}
O_k = \{x\in \mathbb{R}^n:
\left(\begin{array}{c}x\\1 \end{array}\right)^T \bar{Q} \left(\begin{array}{c}x\\1 \end{array}\right) \le 0,  \bar{Q}\in \mathcal{Q}_k
\}.
\end{align}
for all $k\in \mathbb{Z}^+_0$. Since Problem (\ref{eqn:gmax}) is non-convex, we use the S-procedure (see Section 2.6.3 in \citep{BOO:BGFB94} for details) to verify set invariance. More precisely, we check the redundancy of the new quadratic constraints generated in (\ref{eqn:Ok}) by solving a set of LMIs, which is formally stated in the following lemma.

\begin{lemma}\label{lem:sprocedure}
Suppose $\Theta=\mathbb{R}^n$ and $X=\Omega$. Let $O_k$ be defined by the procedure in  (\ref{eqn:O0})-(\ref{eqn:Ok}),  and $\mathcal{Q}_k$ be defined in (\ref{eqn:mathcalQ0})-(\ref{eqn:mathcalQk}) for all $k \in \mathbb{Z}^+_0$. If, for some $k \in \mathbb{Z}^+_0$ and every  $\bar{Q}'  \in  \mathcal{Q}_{k+1} \setminus \mathcal{Q}_{k}$,  there exists $\bar{Q}\in cone(\mathcal{Q}_{k})$  such that 
$
\bar{Q}' \preceq \bar{Q}
$
, then, $O_{k+1} = O_k$.\\
\end{lemma}
Proof of Lemma \ref{lem:sprocedure}: This is a direct application of the S-procedure \citep{BOO:BGFB94}.  Suppose, for every  $\bar{Q}'  \in  \mathcal{Q}_{k+1} \setminus \mathcal{Q}_{k}$, there exists $\bar{Q}\in cone(\mathcal{Q}_{k})$ such that 
$
\bar{Q}' \preceq \bar{Q}
$, the following inequality holds
\begin{align}
\left(\begin{array}{c}x\\1 \end{array}\right)^T \bar{Q}' \left(\begin{array}{c}x\\1 \end{array}\right) 
\preceq \left(\begin{array}{c}x\\1 \end{array}\right)^T \bar{Q}\left(\begin{array}{c}x\\1 \end{array}\right).
\end{align}
for any $x\in \R^{n}$. From (\ref{eqn:OkmathcalQk}), the right hand side of the inequality above is smaller or equal to $0$ for any $x\in O_k$. Hence, $O_k$ is a subset of the set $\Delta O_k := \{x:\left(\begin{array}{c}x\\1 \end{array}\right)^T \bar{Q}'\left(\begin{array}{c}x\\1 \end{array}\right) \le 0,\bar{Q}'\in \mathcal{Q}_{k+1} \setminus \mathcal{Q}_{k}\}$, which implies $O_{k+1} = O_k \cap \Delta O_k = O_k$.
$\Box$

As we have seen, under Assumptions \ref{ass:A} and \ref{ass:interior}, the formal algorithm described in (\ref{eqn:O0})-(\ref{eqn:Ok}) always terminates in finite time. This algorithm is easily implementable when $X$ is a polytope, see \citep{ART:GKT95,ART:Bla99}. In many cases,  it is not directly implementable in the presence of nonlinear constraints. Even if $X$ is convex, the optimization problem (\ref{eqn:gmax}) is still non-convex. However, the same algorithm with the S-procedure in Lemma \ref{lem:sprocedure} is practically implementable, since these LMIs can be efficiently solved using interior point methods \citep{BOO:BGFB94}. To recover the nice finite termination property of the formal algorithm, the following fact is needed.

\begin{fact}\label{fact:Dx}
There exists $D_x > 0$ such that $\|x\|^2\le D_x$ for all $x\in \Omega$.
\end{fact}

This fact always holds under Assumption \ref{ass:interior}. Indeed, without loss of generality, we can always add a redundant ball constraint of the form $\|x\|^2\le D_x$ to $\Omega$. With this fact, we can let $Q_1=\frac{1}{D_x}I$ and $q_1=0$ in (\ref{eqn:omegaQi}). We now show that the finiteness property of the former algorithm in (\ref{eqn:O0})-(\ref{eqn:Ok}) still holds for the LMI version.

\begin{lemma}\label{lem:SpQ}
Suppose Assumptions \ref{ass:A} and \ref{ass:interior} hold, $\Theta=\mathbb{R}^n$, and $X=\Omega$ with $Q_1=\frac{1}{D_x}I$ and $q_1=0$ in (\ref{eqn:omegaQi}). Let $O_k$ be defined by the procedure in  (\ref{eqn:O0})-(\ref{eqn:Ok}) for all $k \in \mathbb{Z}^+_0$. Then, for any $i \in \mathcal{I}_p$, there exists some $k_i\in \mathbb{Z}^+_0$ such that $(\bar{A}^{k_i+1})^T\bar{Q}_i \bar{A}^{k_i+1}\preceq \bar{Q}$ ($\bar{A}$ and $\bar{Q}_i$ are given in (\ref{eqn:barA}) and (\ref{eqn:barQi}) respectively) for some $\bar{Q}\in cone(\mathcal{Q}_{k_i} )$.
\end{lemma}
Proof of Lemma \ref{lem:SpQ}: From (\ref{eqn:barA}), (\ref{eqn:barQi}) and (\ref{eqn:deltaQk}), we have
\begin{align*}
(\bar{A}^{k+1})^T\bar{Q}_i \bar{A}^{k+1} = \left(\begin{array}{cc} (A^{k+1})^TQ_iA^{k+1} & (A^{k+1})^Tq_i\\ q_i^TA^{k+1} & -1 \end{array}\right)
\end{align*}
for all $i\in \mathcal{I}_p$ and $k\in \Z^+_0$. From Assumption \ref{ass:A}, $A^{k}$ goes to $0$ as $k$ increases.  With the additional redundant constraint  $\|x\|^2 \le D_x$, there always exists $Q_j \succ 0$ for some $j\in \mathcal{I}_p$ (one obvious choice is $j=1$), which means that there exists a constant $c>0$ such that 
\begin{align*}
 \left(\begin{array}{cc} Q_j & q_j\\ q_j^T& c \end{array}\right) \succ 0.
\end{align*}
Hence, for any $\beta \in (0, \frac{1}{1+c}]$, 
\begin{align}
\left(\begin{array}{cc} Q_j& q_j\\ q_j^T & \frac{1}{\beta}-1 \end{array}\right) \succeq \left(\begin{array}{cc} Q_j & q_j\\ q_j^T& c \end{array}\right) \succ 0.
\end{align}
From the inequality above and the fact that $A^{k}$ goes to $0$ as $k$ increases, given any $\beta \in (0, \frac{1}{1+c}]$, for any $i\in \mathcal{I}_p$, there always exists a $k_i$ such that 
\begin{align}\label{eqn:QiAtau01}
&(\bar{A}^{k_i+1})^T\bar{Q}_i \bar{A}^{k_i+1}  -  \beta \left(\begin{array}{cc} Q_j& q_j\\ q_j^T & -1 \end{array}\right) \nonumber\\
& =  \left(\begin{array}{cc} (A^{k_i+1})^TQ_iA^{k_i+1}   & (A^{k_i+1})^Tq_i\\ q_i^TA^{k_i+1} & 0 \end{array}\right) \nonumber \\
& ~~~- \beta\left(\begin{array}{cc} Q_j& q_j\\ q_j^T & \frac{1}{\beta}-1 \end{array}\right)  \preceq 0.
\end{align} 
 Therefore, $(\bar{A}^{k_i+1})^T\bar{Q}_i \bar{A}^{k_i+1} \preceq \beta \left(\begin{array}{cc} Q_j& q_j\\ q_j^T & -1 \end{array}\right) \in cone(\mathcal{Q}_{k_i})$.  This completes the proof.
$\Box$

\begin{remark}
As shown in the proof of Lemma \ref{lem:SpQ}, the purpose of adding the redundant constraint $\|x\|^2 \le D_x$ is to provide a guaranteed bound on $(\bar{A}^{k+1})^T\bar{Q}_i \bar{A}^{k+1}$ for all $i\in \mathcal{I}_p$ for sufficiently large $k\in \mathbb{Z}^+_0$. If there already exists $Q_j \succ 0$ for some $j\in \mathcal{I}_p$, it is not necessary to add this constraint.
\end{remark}

Based on Lemma \ref{lem:SpQ}, the following LMI optimization problem is defined for all $Q \in \mathcal{Q}_{k+1}\setminus \mathcal{Q}_{k}$ and $k\in \mathbb{Z}^+_0$:
\begin{subequations}\label{eqn:GAOkr}
\begin{align}
\mathcal{R}(Q,\mathcal{Q}_k ) := &\min\limits_{r,\pmb{\tau}} r\\
s.t. \quad &Q \preceq \sum_{\bar{Q}\in \mathcal{Q}_k}  \tau_{\bar{Q}} \bar{Q}+ rI, \label{eqn:barQir}\\
&  \pmb{\tau}\ge 0, 
\end{align}
\end{subequations}
where $\pmb{\tau}:=\{\tau_{\bar{Q}}, \bar{Q} \in \mathcal{Q}_k\}$. Some properties of the LMI problem above are stated in the following lemma.

\begin{lemma}\label{lem:rquad}
Suppose Assumptions \ref{ass:A} and \ref{ass:interior} hold, $\Theta=\mathbb{R}^n$, and $X=\Omega$. Let $\mathcal{Q}_k$ be defined in (\ref{eqn:mathcalQ0})-(\ref{eqn:mathcalQk}) for all $k \in \mathbb{Z}^+_0$. The optimum of Problem (\ref{eqn:GAOkr}) is denoted by $\mathcal{R}(Q,\mathcal{Q}_k) $ for all $Q \in \mathcal{Q}_{k+1}\setminus \mathcal{Q}_{k}$ and $k\in \mathbb{Z}^+_0$. Then, for any $Q \in \mathcal{Q}_{k+1}\setminus \mathcal{Q}_{k}$, $\mathcal{R}(Q,\mathcal{Q}_k) \le 0$ implies $\mathcal{R}(\bar{A}^TQ\bar{A},\mathcal{Q}_{k+1} ) \le 0$, where $\bar{A}$ is given in (\ref{eqn:barA}).
\end{lemma}
Proof of Lemma \ref{lem:rquad}: Suppose $\mathcal{R}(Q,\mathcal{Q}_k)\le 0$ and the optimal solution is $(\mathcal{R}(Q,\mathcal{Q}_k),\bar{\pmb{\tau}})$, we have 
$
Q \preceq \sum_{\bar{Q}\in \mathcal{Q}_{k}} \bar{\tau}_{\bar{Q}} \bar{Q},
$
which implies that
\begin{align}
\bar{A}^T Q \bar{A}  \preceq \sum_{\bar{Q}\in \mathcal{Q}_{k}} \bar{\tau}_{\bar{Q}} \bar{A}^T \bar{Q} \bar{A} .\label{eqn:barGik1}
\end{align}
As shown below, we can obtain a feasible solution for Problem (\ref{eqn:GAOkr}) with the pair $(\bar{A}^TQ\bar{A},\mathcal{Q}_{k+1})$. Let 
$\pmb{\tau}':=\{\tau'_{\bar{Q}'} , \bar{Q}' \in \mathcal{Q}_{k+1}\}$ be given as follows:
\begin{align}
\tau'_{\bar{Q}'} =  \begin{cases} 
      \bar{\tau}_{\bar{Q}} &  \bar{Q}'\in \bar{A}^T \mathcal{Q}_{k} \bar{A} ,\\
      0 & \bar{Q}' \in \mathcal{Q}_{0},
   \end{cases}
\end{align}
where $\bar{Q}\in \mathcal{Q}_{k}$ is corresponding matrix that satisfies $\bar{Q}'=\bar{A}^T\bar{Q}\bar{A}$ for $\bar{Q}'\in \bar{A}^T \mathcal{Q}_{k} \bar{A}$. Consider that $\mathcal{Q}_{k+1} = \bar{A}^T \mathcal{Q}_{k} \bar{A} \cup \mathcal{Q}_{0}$ for all $k\in \mathbb{Z}^+_0$ from (\ref{eqn:mathcalQ0})-(\ref{eqn:mathcalQk}), (\ref{eqn:barGik1}) impies that $(0,\pmb{\tau}')$ is a feasible solution to Problem (\ref{eqn:GAOkr}) with the pair $(\bar{A}^TQ\bar{A},\mathcal{Q}_{k+1})$ for any  $Q \in \mathcal{Q}_{k+1}\setminus \mathcal{Q}_{k}$ and thus $\mathcal{R}(\bar{A}^TQ\bar{A},\mathcal{Q}_{k+1}) \le 0$. $\Box$

In the following theorem, we show that the LMI problem (\ref{eqn:GAOkr}) can be used to establish a stopping criterion for the algorithm summarized in (\ref{eqn:O0})-(\ref{eqn:Ok}).

\begin{theorem}\label{thm:quad}
Suppose Assumptions \ref{ass:A} and \ref{ass:interior} hold, $\Theta=\mathbb{R}^n$, and $X=\Omega$ with $Q_1=\frac{1}{D_x}I$ and $q_1=0$ in (\ref{eqn:omegaQi}). Let $\mathcal{Q}_k$ be defined in (\ref{eqn:mathcalQ0})-(\ref{eqn:mathcalQk}) for all $k \in \mathbb{Z}^+_0$. For all $k\in \mathbb{Z}^+_0$ and $Q \in \mathcal{Q}_{k+1}\setminus \mathcal{Q}_{k}$, define $\mathcal{R}(Q,\mathcal{Q}_k) $  as in (\ref{eqn:GAOkr}) and let $\mathcal{R}_{\max}(k):=\max_{Q\in \mathcal{Q}_{k+1}\setminus \mathcal{Q}_{k}} \mathcal{R}(Q,\mathcal{Q}_k) $. Then, there exists some finite $k^*$ such that $\mathcal{R}_{\max}(k^*)\le 0$ and $O_{\infty} = O_{k^*}$.
\end{theorem}
Proof of Theorem \ref{thm:quad}: From Lemmas \ref{lem:SpQ} and \ref{lem:rquad}, there always exists some $k_i$ such that  $\mathcal{R}((\bar{A}^{k+1})^TQ_i \bar{A}^{k+1},\mathcal{Q}_{k})  \le 0$  for all $k \ge k_i$ and $i \in \mathcal{I}_p$. Let $k^*:=\max_{i \in \mathcal{I}_p} k_i$. We can see that $\mathcal{R}((\bar{A}^{k^*+1})^TQ_i \bar{A}^{k^*+1},\mathcal{Q}_{k^*})\le 0$ for all $i \in \mathcal{I}_p$, which implies $\mathcal{R}_{\max}(k^*)\le 0$. Following Lemma \ref{lem:sprocedure}, we can get $O_{k^*+1} = O_{k^*}$. Finally, it holds that $O_{\infty} = O_{k^*}$.
$\Box$

From Theorem \ref{thm:quad}, the maximal \emph{CA-invariant} set $O_{\infty}$ can be exactly characterized by $\{\mathcal{R}_{\max}(k)\}_{k\in \mathbb{Z}_0^+}$ with guaranteed finite determination. The determination condition ($\mathcal{R}_{\max}(k) \le 0$ for some $k\in \mathbb{Z}_0^+$) is computationally tractable and leads to the true $O_{\infty}$.

Based on the discussion above, the algorithm to compute the maximal \emph{CA-invariant} set with quadratic constraints is summarized in Algorithm \ref{algo:quad}.

\begin{algorithm}[h]
\caption{Computation of the maximal \emph{CA-invariant} set with quadratic constraints}
\hspace*{\algorithmicindent} \textbf{Input}: $A$ and $\{Q_i,q_i\}_{i=1}^p$ as in (\ref{eqn:omegaQi})\\
\hspace*{\algorithmicindent} \textbf{Output}: $O_{k^*}$
\begin{algorithmic}[1]
\STATE \textit{Initialization}: let $X:=\{x\in \mathbb{R}^n : x^TQ_ix +2q_i^Tx\le 1, i\in \mathcal{I}_p\}$, set $k=0$ and $O_0 = X$, and construct $\mathcal{Q}_0$ as in (\ref{eqn:mathcalQ0});
\STATE  Let $\mathcal{Q}_{k+1}$ be updated according to (\ref{eqn:mathcalQk});
\STATE Obtain $\mathcal{R}(Q,\mathcal{Q}_k )$ from (\ref{eqn:GAOkr}) for all $Q \in \mathcal{Q}_{k+1}\setminus \mathcal{Q}_k$;
\STATE Let $\mathcal{R}_{\max}(k):=\max_{Q\in \mathcal{Q}_{k+1}\setminus \mathcal{Q}_{k}} \mathcal{R}(Q,\mathcal{Q}_k)$. If $\mathcal{R}_{\max}(k) \le 0$, let $k^*=k$ and terminate; otherwise, let $O_{k+1} := O_k \bigcap \{x\in \mathbb{R}^n: Ax\in O_k\}$, set $k \leftarrow k+1$ and go to Step 2.
\end{algorithmic}
\label{algo:quad}
\end{algorithm}

Since $|\mathcal{Q}_{k}| \le  (k+1)p$ and $|\mathcal{Q}_{k+1}\setminus\mathcal{Q}_{k}|  \le  p$, $k\in \mathbb{Z}^+_0$, at the $k^{th}$ iteration in Algorithm \ref{algo:quad}, we solve  at most  $p$ LMI problems with  at most  $(k+1)p+1$ variables and one LMI constraint. As $k$ increases, $\mathcal{Q}_k$ may have some redundant elements, which can be removed using a similar formulation as (\ref{eqn:GAOkr}): 
\begin{subequations}\label{eqn:Okred}
\begin{align}
\mathcal{R}(Q,\mathcal{Q}_k\setminus Q ) = &\min\limits_{r,\pmb{\tau}} r\\
s.t. \quad & Q \preceq \sum_{\bar{Q}\in \mathcal{Q}_k\setminus Q}  \tau_{\bar{Q}} \bar{Q}+ rI,\\
& \pmb{\tau}\ge 0, 
\end{align}
\end{subequations}
where $\pmb{\tau}=\{\tau_{\bar{Q}}: \bar{Q} \in \mathcal{Q}_k\setminus Q \}$ for any $Q \in \mathcal{Q}_k$. If, for some $Q \in \mathcal{Q}_k$ at the $k^{th}$ iteration, $\mathcal{R}(Q,\mathcal{Q}_k\setminus Q )\le 0$, then, $Q$ is redundant and can be removed from $\mathcal{Q}_k$. After all the redundant elements are removed, a reduced set of $\mathcal{Q}_k$ can be obtained. Since removing redundant elements from $\mathcal{Q}_k$ does not change the sign of the optimum of Problem (\ref{eqn:GAOkr}), the results in Theorem \ref{thm:quad} are still valid. 

As (\ref{eqn:gmax}) is not directly solved, the $k^*$ obtained from Algorithm \ref{algo:quad} is an upper bound on $k_{\min}$. For a loose upper bound $k^*$, the description of $O_{k^*}$ may not be tight enough though it is still true that $O_{k^*}=O_{\infty}$. However, in some cases, $k^*$ is not necessarily a loose upper bound. It can be close or equal to $k_{\min}$. One example is the case with only linear constraints, i.e., $\Theta=\mathbb{R}^n$ and $Q_i=0$ for all $i\in \mathcal{I}_p$. The proposition below shows that the $k^*$ obtained from Algorithm \ref{algo:quad} is exactly equal to $k_{\min}$ in the case of linear constraints. 

\begin{proposition}\label{lem:linear}
Suppose Assumptions \ref{ass:A} and \ref{ass:interior} hold, $\Theta=\mathbb{R}^n$ and $Q_i=0$ for all $i\in \mathcal{I}_p$. The constraint set $X$ can be expressed as $\{x\in \mathbb{R}^n:  2q^Tx \le \pmb{1}_p\}$, where $q:= [q_1~q_2~\cdots ~q_p]$. For any $k\in \mathbb{Z}_0^+$, let $\mathcal{R}_{\max}(k)$ and $O_k$ be generated by Algorithm \ref{algo:quad}. Then, it holds that $\mathcal{R}_{\max}(k)\le 0$ if and only if $O_{k+1} = O_k$ for $k\in \mathbb{Z}_0^+$.
\end{proposition}

The proof of Proposition \ref{lem:linear} is given in the appendix. From Proposition \ref{lem:linear}, we can see that Algorithm \ref{algo:quad} is eventually equivalent to the standard algorithm \citep{ART:Bla99} for linear systems with linear constraints. Generally speaking, the conservatism of $k^*$ obtained from Algorithm \ref{algo:quad} depends on the conservatism of the S-procedure in Lemma \ref{lem:sprocedure}. If the LMI in Lemma \ref{lem:sprocedure} is a necessary and sufficient condition of the set inclusion in (\ref{eqn:OkAX}), the S-procedure is lossless and $k^*$ is exactly equal to $k_{\min}$. However, for general quadratic constraints, this is not true. A detailed discussion on the conservatism of S-procedure can be found in \citep{ART:DP06}. More precisely, $k^*$ can be larger than $k_{\min}$ in most of the cases. However, the size of the resulting $O_{\infty}$ is not affected although there are redundant constraints in the description of the set. With Fact \ref{fact:Dx}, another possibility to determine a $k$ that satisfies (\ref{eqn:OkAX}) is to find a $k$ such that $A^{k+1}x$ enters an open ball inside $X$ for any $x\in \{x:\|x\|^2\le D_x\}$. However, this is usually very conservative and such a $k$ can be much larger than the $k^*$ obtained from Algorithm \ref{algo:quad}.

\subsection{Quasi-smooth nonlinear constraints}\label{sec:quadnl}
In the rest of this section, the proposed approach will be generalized to handle non-quadratic nonlinear constraints that satisfy Assumption \ref{ass:smooth}. This is possible by making use of
the quadratic upper and lower bounds in (\ref{eqn:Hi0x}). With these quadratic bounds, we are able to establish quadratic relaxations of (\ref{eqn:gmax}) and (\ref{eqn:hmax}).  More precisely, the constraints in (\ref{eqn:gmax}) and (\ref{eqn:hmax}) are replaced by their quadratic lower bounds and the objectives in (\ref{eqn:hmax}) are replaced by their quadratic upper bounds.  For notational simplicity, let
\begin{align}
H^u_i(x) :&= H_i(0) + (H_{i}^{\nabla})^T x + \frac{L_i}{2} \|x\|^2 \nonumber\\
&=\left(\begin{array}{c}
x\\
1
\end{array}\right)^T\left(\begin{array}{cc} \frac{L_i}{2}I & \frac{1}{2}H_{i}^{\nabla} \\ \frac{1}{2}(H_{i}^{\nabla})^T & H_i(0) \end{array}\right) \left(\begin{array}{c}
x\\
1
\end{array}\right), \label{eqn:Hupper}\\
H^l_i(x) : &= H_i(0) + (H_{i}^{\nabla})^T x - \frac{L_i}{2} \|x\|^2 \nonumber\\
&=\left(\begin{array}{c}
x\\
1
\end{array}\right)^T\left(\begin{array}{cc} -\frac{L_i}{2}I & \frac{1}{2}H_{i}^{\nabla} \\ \frac{1}{2}(H_{i}^{\nabla})^T & H_i(0) \end{array}\right)\left(\begin{array}{c}
x\\
1
\end{array}\right), \label{eqn:hatHi}
\end{align}
for all $i\in \mathcal{I}_m$. Similar to (\ref{eqn:mathcalQ0})-(\ref{eqn:mathcalQk}), we define:
\begin{align}
\mathcal{H}^u_0 &= \{\bar{H}^u_i,  i\in \mathcal{I}_m\}, \label{eqn:Hu0}\\
\mathcal{H}^l_0 &= \{\bar{H}^l_i, i\in \mathcal{I}_m\}, \label{eqn:Hl0}\\
\mathcal{H}^u_{k+1} &:= \{\bar{A} ^T\bar{Q} \bar{A}: \bar{Q}\in \mathcal{H}^u_{k}\} ,\label{eqn:Huk}\\
\mathcal{H}^l_{k+1} &:= \mathcal{H}^l_{k} \bigcup \{\bar{A} ^T\bar{Q} \bar{A}: \bar{Q}\in \mathcal{H}^l_{k}\}, \label{eqn:Hlk}
\end{align}
where 
\begin{align}
\bar{H}^u_i &= \left(\begin{array}{cc} \frac{L_i}{2}I & \frac{1}{2}H_{i}^{\nabla} \\ \frac{1}{2}(H_{i}^{\nabla})^T & H_i(0)-1 \end{array}\right), \label{eqn:barHui}\\
\bar{H}^l_i &=  \left(\begin{array}{cc} -\frac{L_i}{2}I & \frac{1}{2}H_{i}^{\nabla} \\ \frac{1}{2}(H_{i}^{\nabla})^T & H_i(0)-1 \end{array}\right). \label{eqn:barHli}
\end{align}
The sets $\{\mathcal{H}^u_k\}$ and $\{\mathcal{H}^l_k\}$ are updated differently because $\{\mathcal{H}^u_k\}$ is used in the cost function while $\{\mathcal{H}^l_k\}$ is used in the constraints as shown later. With additional definitions above, a relaxed quadratic constraint set of $O_k$ can be obtained for all $k\in \mathbb{Z}_0^+$:
\begin{align}
\tilde{O}_k := \{x: &\left(\begin{array}{c}x\\1 \end{array}\right)^T\bar{Q}  \left(\begin{array}{c}x\\1 \end{array}\right) \le 0, \bar{Q}\in \mathcal{Q}_k \cup \mathcal{H}^l_{k} \}. \label{eqn:tildeOk}
\end{align}
Based on this relaxed constraint set, a modification of (\ref{eqn:gmax}) is given by
\begin{subequations}\label{eqn:gmaxbar}
\begin{align}
\bar{g}_i(k) : = &\max\limits_{x} (A^{k+1}x)^TQ_i A^{k+1}x+2q_i^TA^{k+1}x\\
\textrm{s.t.} \quad & x \in \tilde{O}_k,
\end{align}
\end{subequations}
for any $i \in \mathcal{I}_p$ and $k\in \mathbb{Z}_0^+$. As $O_k \subseteq \tilde{O}_k$, $\bar{g}_i(k) \ge g_i(k)$ for all $i \in \mathcal{I}_p$ and $k\in \mathbb{Z}_0^+$. Similarly, we can also modify (\ref{eqn:hmax}) using the relaxed set. Since the cost function of (\ref{eqn:hmax}) is also nonlinear, we will replace it by its quadratic upper bound (\ref{eqn:Hupper}). With the relaxed set and the quadratic upper bound of the cost function, the corresponding modification of (\ref{eqn:hmax}) is given by
\begin{subequations}\label{eqn:hmaxbarmqm}
\begin{align}
\bar{h}_i(k) : = &\max\limits_{x} H^u_i(A^{k+1}x)\\
\textrm{s.t.} \quad & x \in \tilde{O}_k
\end{align}
\end{subequations}
for all $i\in \mathcal{I}_m$. Again, we can see that $\bar{h}_i(k)\ge h_i(k)$ for all $i\in \mathcal{I}_m$ and $k\in \mathbb{Z}_0^+$. Using the S-procedure, the following lemma can be obtained immediately.

\begin{lemma}\label{lem:Spnl}
Suppose Assumption \ref{ass:smooth} holds. Let the set $O_k$ be defined by the procedure in  (\ref{eqn:O0})-(\ref{eqn:Ok}) and the relaxed quadratic set $\tilde{O}_k$ be defined in (\ref{eqn:tildeOk}) using the quadratic lower bounds (\ref{eqn:hatHi}) for all $k\in \mathbb{Z}^+_0$. Consider the sets $\{\mathcal{Q}_k,\mathcal{H}^u_k,\mathcal{H}^l_k\}$ defined in (\ref{eqn:mathcalQ0})-(\ref{eqn:mathcalQk}) and (\ref{eqn:Hu0})-(\ref{eqn:Hlk}), the following results hold.\\
(i) For any $i\in \mathcal{I}_p$, if $
(\bar{A}^{k+1})^T \bar{Q}_i\bar{A}^{k+1} \preceq \bar{Q}
$ ($\bar{A}$ and $\bar{Q}_i$ are given in (\ref{eqn:barA}) and (\ref{eqn:barQi}) respectively) is satisfied for some $\bar{Q} \in cone(\mathcal{Q}_k\cup \mathcal{H}^l_k)$ and some $k\in \mathbb{Z}_0^+$, then, 
\begin{align}
\left(\begin{array}{c}x\\1 \end{array}\right)^T (\bar{A}^{k+1})^T \bar{Q}_i\bar{A}^{k+1}  \left(\begin{array}{c}x\\1 \end{array}\right) \le 0, \forall x\in O_k.
\end{align}
(ii) For any $i\in \mathcal{I}_m$, if $
(\bar{A}^{k+1})^T \bar{H}_i^u\bar{A}^{k+1}  \preceq \bar{Q}
$ ($\bar{H}_i^u$ is given in (\ref{eqn:barHui})) is satisfied for some $\bar{Q} \in cone(\mathcal{Q}_k\cup \mathcal{H}^l_k)$ and some $k\in \mathbb{Z}_0^+$, then, $H_i(A^{k+1}x)\le 1$ for all $x\in O_k$. 
\end{lemma}
Proof of Lemma \ref{lem:Spnl}: (i) An immediate consequence of the S-procedure is that  $\left(\begin{array}{c}x\\1 \end{array}\right)^T \bar{Q} \left(\begin{array}{c}x\\1 \end{array}\right) \le 0$  for any $x\in \tilde{O}_k$. Taking into account that $O_k\subseteq \tilde{O}_k$, property (i) holds true.\\
(ii) Similarly, from the S-procedure, $H^u_i(A^{k+1}x)\le 1$ for any $x\in \tilde{O}_k$. Since $H_i(A^{k+1}x)\le H^u_i(A^{k+1}x)$ for any $x\in \Omega$ and $O_k\subseteq \tilde{O}_k$, property (ii) is proved.
$\Box$

From the lemma above, we can see that it is also possible to implement the formal algorithm in (\ref{eqn:O0})-(\ref{eqn:Ok}) using the quadratic relaxations in (\ref{eqn:Hupper})-(\ref {eqn:hatHi}) for general nonlinear constraints that satisfy Assumption \ref{ass:smooth}. The finite termination of the algorithm is discussed in the next lemma.

\begin{lemma}\label{lem:Spnlfinite}
Suppose Assumptions \ref{ass:A}-\ref{ass:smooth} hold  with $Q_1=\frac{1}{D_x}I$ and $q_1=0$ in (\ref{eqn:omegaQi}).  Consider the relaxed quadratic set $\tilde{O}_k$ defined in (\ref{eqn:tildeOk}) using the quadratic lower bounds (\ref{eqn:hatHi}), and the sets $\{\mathcal{Q}_k,\mathcal{H}^u_k,\mathcal{H}^l_k\}$ defined in (\ref{eqn:mathcalQ0})-(\ref{eqn:mathcalQk}) and (\ref{eqn:Hu0})-(\ref{eqn:Hlk}) for all $k\in \mathbb{Z}^+_0$, the following results hold.\\
(i) For any $i \in \mathcal{I}_p$, there exists some finite $k_i$ such that $
(\bar{A}^{k_i+1})^T \bar{Q}_i\bar{A}^{k_i+1} \preceq \bar{Q}
$ ($\bar{A}$ and $\bar{Q}_i$ are given in (\ref{eqn:barA}) and (\ref{eqn:barQi}) respectively)
for some $\bar{Q} \in  cone(\mathcal{Q}_{k_i}\cup \mathcal{H}^l_{k_i})$.\\
(ii) For any $i\in \mathcal{I}_m$,  there exists some finite $k_i$ such that
$
(\bar{A}^{k_i+1})^T \bar{H}^u_i\bar{A}^{k_i+1}  \preceq \bar{Q}
$ ($\bar{H}_i^u$ is given in (\ref{eqn:barHui}))
for some $\bar{Q} \in  cone(\mathcal{Q}_{k_i}\cup \mathcal{H}^l_{k_i})$.
\end{lemma}
Proof of Lemma \ref{lem:Spnlfinite}: The proof follows the same arguments in Lemma \ref{lem:SpQ} and thus is omitted. $\Box$

Based on Lemma \ref{lem:Spnlfinite}, Problem (\ref{eqn:GAOkr}) is modified as
\begin{subequations} \label{eqn:QQdGHr}
\begin{align}
\mathcal{R}(Q,\mathcal{Q}_k\cup \mathcal{H}^l_k ) := &\min\limits_{r,\pmb{\tau}} r\\
s.t. \quad &Q \preceq \sum_{\bar{Q}\in \mathcal{Q}_k\cup \mathcal{H}^l_k}  \tau_{\bar{Q}} \bar{Q}+ rI, \\
& \pmb{\tau}\ge 0,
\end{align}
\end{subequations}
 for any $Q\in (\mathcal{Q}_{k+1}\setminus \mathcal{Q}_k) \cup \mathcal{H}^u_{k+1}$ and $k\in \mathbb{Z}^+_0$. The following lemma can be derived.

\begin{lemma}\label{lem:rtilder}
Suppose Assumptions \ref{ass:A}-\ref{ass:smooth} hold. Let the sets $\{\mathcal{Q}_k,\mathcal{H}^u_k,\mathcal{H}^l_k\}$ be defined in (\ref{eqn:mathcalQ0})-(\ref{eqn:mathcalQk}) and (\ref{eqn:Hu0})-(\ref{eqn:Hlk}) for all $k\in \mathbb{Z}^+_0$. Let $\mathcal{R}(Q,\mathcal{Q}_k\cup \mathcal{H}^l_k ) $ be defined in (\ref{eqn:QQdGHr}) for any $Q\in (\mathcal{Q}_{k+1}\setminus \mathcal{Q}_k) \cup \mathcal{H}^u_{k+1}$ and $k\in \mathbb{Z}^+_0$. The following properties hold.\\
(i) For any $i \in \mathcal{I}_p$, there exists a finite $k_i\in \mathbb{Z}_0^+$ such that $\mathcal{R}((\bar{A}^{k+1})^T \bar{Q}_i(\bar{A}^{k+1} ,\mathcal{Q}_k\cup \mathcal{H}^l_k )\le 0$ ($\bar{A}$ and $\bar{Q}_i$ are given in (\ref{eqn:barA}) and (\ref{eqn:barQi}) respectively) for all $k\ge k_i$.\\
(ii) For all $i\in \mathcal{I}_m$, there exists a finite $k_i\in \mathbb{Z}_0^+$ such that $\mathcal{R}((\bar{A}^{k+1})^T
\bar{H}_i^u\bar{A}^{k+1}, \mathcal{Q}_{k}\cup \mathcal{H}^l_{k})\le 0
$ ($\bar{A}$ and $\bar{H}_i^u$ are given in (\ref{eqn:barA}) and (\ref{eqn:barHui}) respectively)
for all $k\ge k_i$.
\end{lemma}
Proof of Lemma \ref{lem:rtilder}: The proof follows the same arguments in Lemma \ref{lem:rquad} and hence is omitted.
$\Box$

Based on Lemmas \ref{lem:Spnl} - \ref{lem:rtilder}, the algorithm for computing the maximal \emph{CA-invariant} set with nonlinear constraints is summarized in Algorithm \ref{algo:nl}. At each iteration $k$ of Algorithm \ref{algo:quad} for $k\in \mathbb{Z}^+_0$, we solve  at most  $p+m$ LMI problems with  at most  $(k+1)(p+m)+1$ variables and one LMI constraint. Similar to Algorithm \ref{algo:quad}, Algorithm \ref{algo:nl} will also terminate after a finite time as stated in Theorem \ref{thm:quadnl}.

\begin{algorithm}[h]
\caption{Computation of the maximal constraint admissible invariant set with nonlinear constraints}
\hspace*{\algorithmicindent} \textbf{Input}: $A$, $\{Q_i,q_i\}_{i=1}^{p}$, and $\{H_i(x),H_{i}^{\nabla},L_i\}_{i=1}^m$\\
\hspace*{\algorithmicindent} \textbf{Output}: $O_{k^*}$
\begin{algorithmic}[1]
\STATE \textit{Initialization}: let $X:=\{x\in \mathbb{R}^n:  (x)^TQ_ix+2q^T_ix\le 1, i \in \mathcal{I}_p, H(x)\le 0\}$, set $k=0$ and $O_0=X$, construct $\mathcal{Q}_0$, $\mathcal{H}^u_0$ and $\mathcal{H}^l_0$ as in (\ref{eqn:mathcalQ0}), (\ref{eqn:Hu0}) and (\ref{eqn:Hl0}) respectively;
\STATE Update $\mathcal{Q}_{k+1}$, $\mathcal{H}^u_{k+1}$ and $\mathcal{H}^l_{k+1}$ according to (\ref{eqn:mathcalQk}), (\ref{eqn:Huk}) and (\ref{eqn:Hlk}) respectively;
\STATE  Obtain $\mathcal{R}(Q,\mathcal{Q}_k\cup \mathcal{H}^l_k ) $ for any $Q\in (\mathcal{Q}_{k+1}\setminus \mathcal{Q}_k) \cup \mathcal{H}^u_{k+1}$;
\STATE Let $\mathcal{R}_{\max}(k):=\max_{Q\in (\mathcal{Q}_{k+1}\setminus \mathcal{Q}_k) \cup \mathcal{H}^u_{k+1}}\mathcal{R}(Q,\mathcal{Q}_k\cup \mathcal{H}^l_k ) $. If $\mathcal{R}_{\max}(k) \le 0$, let $k^*=k$ and terminate; otherwise, let $O_{k+1} := O_k \bigcap \{x\in \mathbb{R}^n: Ax\in O_k\}$, set $k \leftarrow k+1$ and go to Step 2.
\end{algorithmic}
\label{algo:nl}
\end{algorithm}

\begin{theorem}\label{thm:quadnl}
Suppose Assumptions \ref{ass:A}-\ref{ass:smooth} hold  with $Q_1=\frac{1}{D_x}I$ and $q_1=0$ in (\ref{eqn:omegaQi}), let $\mathcal{R}_{\max}(k)$ and $O_k$ be generated from Algorithm \ref{algo:nl} for $k\in \mathbb{Z}_0^+$. Then, there exists some finite $k^*$ such that $\mathcal{R}_{\max}(k^*)\le 0$ and $O_{\infty} = O_{k^*}$.
\end{theorem}
Proof of Theorem \ref{thm:quadnl}: The proof follows similar arguments in the proof of Theorem \ref{thm:quad}.
$\Box$

\subsection{Semi-algebraic constraints}
We now consider one special case in which $\Theta$ is a semi-algebraic constraint set and $\{H_i(x)\}_{i=1}^m$ are polynomial functions of degree smaller or equal to $d$. Since quadratic constraints are handled separately, we assume that $d\ge 3$.  Clearly, semi-algebraic constraints satisfy Assumption \ref{ass:smooth} with $H_{i}^{\nabla}=\nabla H_i(0)$ and $L_i$ being the Lipschitz constant in $\Omega$ for all $i\in \mathcal{I}_m$. Although semi-algebraic constraints can be handled by Algorithm \ref{algo:nl}, the Lipschtiz constants $\{L_i\}_{i=1}^m$ can be conservative for high-order polynomial functions. For this reason, we present an alternative method for handling semi-algebraic constraints. In \citep{INP:AJ16}, a lifting method is used to convert semi-algebraic constraints into linear constraints. In this paper, we use a similar lifting method that converts semi-algebraic constraints into quadratic constraints. For the same degree $d$, the dimension of the lifted space in our method can be shown to be lower than the one used in \citep{INP:AJ16}.

The lifting method is described as follows. For any $x\in \mathbb{R}^n$ and $i\in \mathbb{Z}^+$, let $x^{[i]}\in \mathbb{R}^{\binom{n+i-1}{i}}$ denote the vector of all the monomials of degree $i$ and $A^{[i]}:x^{[i]}\rightarrow (Ax)^{[i]}$ denote the lifted linear map of the system (\ref{eqn:xA}). In \citep{INP:AJ16}, semi-algebraic constraints are converted into linear constraints by using this lifted linear map. Thanks to Algorithm \ref{algo:quad}, we only need to convert semi-algebraic constraints into quadratic constraints. This reduces the dimension of the lifted space significantly. With a vector of monomials, the polynomial functions $\{H_i(x)\}_{i=1}^m$ can be always rewritten into quadratic forms, i.e.,
\begin{align}\label{eqn:Hixd}
H_i(x) = \left( \begin{array}{c}
x^{[1]}\\
x^{[2]}\\
\vdots\\
x^{[\bar{d}]}
\end{array} \right)^TP_i \left( \begin{array}{c}
x^{[1]}\\
x^{[2]}\\
\vdots\\
x^{[\bar{d}]}
\end{array} \right) + 2F_i^T x
\end{align}
where 
$\bar{d}= \textrm{ceil}( d/2 )$, $P_i\in \mathbb{R}^{N\times N}$ and $F_i\in \mathbb{R}^{n}$ with $N = \sum_{\ell=1}^{\bar{d}} \binom{n+\ell-1}{\ell}$. The lifted system becomes
\begin{align}\label{eqn:zAtilde}
z(t+1) = \tilde{A}z(t), \quad t \in \Z_0^+
\end{align}
where $z\in \mathbb{R}^N$ and $\tilde{A} = \textrm{diag}\{A^{[1]},A^{[2]},\cdots, A^{[\bar{d}]}\} \in \mathbb{R}^{N\times N}$. From \citep{ART:BN05,INP:AJ16}, $\tilde{A}$ is also Schur stable if $A$ is Schur stable.

The expression in (\ref{eqn:Hixd}) may not be unique and we may only need a subset of $\{x^{[1]},x^{[2]},\cdots, x^{[\bar{d}]}\}$, depending on the polynomial functions. The dimension of the lifted system is $\binom{n+\bar{d}-1}{\bar{d}}$ in the best case (when only $\{x^{[\bar{d}]}\}$ is used) and $\sum_{\ell=1}^{\bar{d}} \binom{n+\ell-1}{\ell}$ in the worst case (when the whole set $\{x^{[1]},x^{[2]},\cdots, x^{[\bar{d}]}\}$ is used). In \citep{INP:AJ16}, the lower and upper bounds  are $\binom{n+d-1}{d}$ and $\sum_{\ell=1}^{d} \binom{n+\ell-1}{\ell}$ respectively. As $\bar{d}= \textrm{ceil}( d/2 )$, the quadratic expression in (\ref{eqn:Hixd}) allows us to significantly reduce the dimension of the lifted space. In fact, it can be verified that our upper bound $\sum_{\ell=1}^{\bar{d}} \binom{n+\ell-1}{\ell}$ is even much smaller than the lower bound $\binom{n+d-1}{d}$ in \citep{INP:AJ16} when $n>2$.

%

In the rest of this section, for ease of discussion and notational simplicity, we consider the whole vector $(x^{[1]},x^{[2]},\cdots, x^{[\bar{d}]})$. As a result, the original quadratic constraints  in (\ref{eqn:omegaQi}) can be expressed as
\begin{align}\label{eqn:zQip}
z^T [I_n ~ \pmb{0}]^TQ_i [I_n ~ \pmb{0}]z+2q_i^T[I_n ~ \pmb{0}] z \le 1, ~~ i \in \mathcal{I}_p,
\end{align}
and the semi-algebraic constraints in (\ref{eqn:xtheta}) become
\begin{align}\label{eqn:zPim}
z^T P_i z + 2F_i^T [I_n ~ \pmb{0}] z\le 1, \quad i\in \mathcal{I}_m. 
\end{align}
Since $\Omega$ is bounded under Assumption \ref{ass:interior}, without loss of generality, we can always add the redundant constraint of the form $\|z\|^2 \le D_z$ for some sufficiently large $D_z>0$ such that $\|(x^{[1]},x^{[2]},\cdots, x^{[\bar{d}]})\|^2\le D_z$ for all $x\in \Omega$. Hence, the overall constraint set of the lifted system can be expressed as
\begin{align}\label{eqn:Xz}
X_z:=\{z\in \mathbb{R}^N: (\ref{eqn:zQip}), ~ (\ref{eqn:zPim}), \textrm{ and } \frac{1}{D_z}\|z\|^2 \le 1\}.
\end{align}
From the definition of $X_z$, it can be verified that $X=\{x\in \mathbb{R}^n: (x^{[1]},x^{[2]},\cdots, x^{[\bar{d}]}) \in X_z\}$. Now, all the constraints in $X_z$ for the lifted system are quadratic (or linear) and we can use Algorithm \ref{algo:quad} to compute the maximal \emph{CA-invariant}  set of the lifted system, denoted by $O_{\infty}^z$. Since $\tilde{A}$ in (\ref{eqn:zAtilde}) is Schur stable, the results in Theorem \ref{thm:quad} are also valid for the lifted system. The following proposition shows that the maximal \emph{CA-invariant}  set of the original system can be exactly characterized by $O_{\infty}^z$.

\begin{proposition}\label{prop:lift}
Suppose Assumptions \ref{ass:A} and \ref{ass:interior} hold and $\{H_i(x)\}_{i=1}^m$ are polynomial functions of degree smaller or equal to $d$. Let $O_{\infty}$ be the maximal \emph{CA-invariant} set of the system (\ref{eqn:xA}) with the constraint set $X$ in (\ref{eqn:xXt}) and $O_{\infty}^z$ be the maximal \emph{CA-invariant} set of the lifted system (\ref{eqn:zQip}) with the constraint set $X_z$ in (\ref{eqn:Xz}). Then,
$
O_{\infty} = \{x\in \mathbb{R}^n: (x^{[1]},x^{[2]},\cdots, x^{[\bar{d}]}) \in O_{\infty}^z\}
$.
\end{proposition}
Proof of Proposition \ref{prop:lift}: First, we show that $O_{\infty}  \subseteq \{x\in \mathbb{R}^n: (x^{[1]},x^{[2]},\cdots, x^{[\bar{d}]}) \in O_{\infty}^z\}$. For any $x\in O_{\infty} $, we know that $A^kx\in X$ for all $k\in \mathbb{Z}^+_0$. From the definition of the lifted system in (\ref{eqn:zAtilde}) and $X_z$ in (\ref{eqn:Xz}), $((A^kx)^{[1]},(A^kx)^{[2]},\cdots, (A^kx)^{[\bar{d}]})=\tilde{A}^k(x^{[1]},x^{[2]},\cdots, x^{[\bar{d}]})\in X_z$ for all $k\in \mathbb{Z}^+_0$, which implies that $(x^{[1]},x^{[2]},\cdots, x^{[\bar{d}]})\in O_{\infty}^z$. Then, we show that $ \{x\in \mathbb{R}^n: (x^{[1]},x^{[2]},\cdots, x^{[\bar{d}]}) \in O_{\infty}^z\} \subseteq O_{\infty}$. For any $x\in \{x\in \mathbb{R}^n: (x^{[1]},x^{[2]},\cdots, x^{[\bar{d}]}) \in O_{\infty}^z\}$, $((A^kx)^{[1]},(A^kx)^{[2]},\cdots, (A^kx)^{[\bar{d}]})\in X_z$ for all $k\in \mathbb{Z}^+_0$. Hence, $A^kx\in X$ for all $k\in \mathbb{Z}^+_0$, which implies that $x\in O_{\infty}$. This completes the proof. $\Box$

It is worth mentioning that the lifting method for semi-algebraic constraints is closely related to sum of squares (SOS) optimization techniques (see, e.g.,  \citep{ART:PW98, ART:P03}  for details of SOS optimization).  More precisely, for the verification of the set inclusion condition (\ref{eqn:OkAX}), from the discussion above, the constraints of affine combinations of quadratic forms that are nonnegative (or nonpositive) for the lifted system are equivalent to the constraints of affine combinations of polynomials that are SOS for the original system.

\section{Particular nonlinear systems}\label{sec:nlsys}
In this section, we show that the proposed approach is also applicable to special types of nonlinear systems.

\subsection{Switched linear systems}
We consider switched linear systems, which are a well-known family of hybrid systems in the form of:
\begin{align}\label{eqn:Asigma}
x(t+1) = A_{\sigma(t)}x(t)
\end{align}
where $\sigma(t): \mathbb{Z}^+\rightarrow \mathcal{I}_M$ is a time-dependent switching signal that indicates the current active mode of the system among $M$ possible modes in $\mathcal{A}:=\{A_1,A_2,\cdots, A_M\}$. For arbitrarily switching systems, the joint spectral radius (JSR) is defined by \citep{BOO:J09}
\begin{align}
\rho(\mathcal{A}): = \lim\limits_{k\rightarrow \infty} \max\limits_{i_1,\cdots, i_k}\{\|A_{i_1}\cdots A_{i_k}\|^{1/k}: A_{i_j}\in \mathcal{A} \}.
\end{align}
As shown in \citep{BOO:J09}, System (\ref{eqn:Asigma}) is asymptotically stable at origin under arbitrary switching if and only if $\rho(\mathcal{A})< 1$. The set invariance of arbitrarily switched linear systems is defined as follows.

\begin{definition}
Given the constraint set $X$ in (\ref{eqn:xXt}), the nonempty set $Z\subseteq X$ is a \emph{CA-invariant}  set for System (\ref{eqn:Asigma}) if $x\in Z$ implies that $A_ix\in Z$ for any $i\in \mathcal{I}_M$.
\end{definition}

As shown in \citep{ART:DO12,INP:AJ16}, the maximal \emph{CA-invariant}  set of System (\ref{eqn:Asigma}) exists if $\rho(\mathcal{A})< 1$ and Assumptions \ref{ass:interior} and \ref{ass:smooth} hold. For its computation, we need to adjust the procedure in (\ref{eqn:O0})-(\ref{eqn:Ok}) as follows:
\begin{align}
O_0 &:= X \label{eqn:Osigma0}\\
O_{k+1} &:= O_k \bigcap \{x: Ax\in O_k, A\in \mathcal{A}\}, k \in \mathbb{Z}^+_0 . \label{eqn:Osigmak}
\end{align}
Let $\bar{\mathcal{A}}:=\{\bar{A}_1,\bar{A}_2,\cdots,\bar{A}_M\}$ with 
\begin{align}
\bar{A}_i &= \left(\begin{array}{cc} A_i & 0\\ 0 & 1 \end{array}\right), \forall i\in \mathcal{I}_M.
\end{align}
The update in (\ref{eqn:mathcalQ0})-(\ref{eqn:mathcalQk}) becomes
\begin{align}
\mathcal{Q}_0 &:= \{\bar{Q}_i, i\in \mathcal{I}_p\} \label{eqn:mathcalQsigma0}\\
\mathcal{Q}_{k+1} &:= \mathcal{Q}_k \bigcup \{\bar{A} ^T\bar{Q} \bar{A}: \bar{Q}\in \mathcal{Q}_k, \bar{A}\in \bar{\mathcal{A}}\},  \label{eqn:mathcalQsigmak}
\end{align}
with $|\mathcal{Q}_{k}|  \le  \sum_{\ell=0}^k M^{\ell}p$, 
\begin{align}
\mathcal{Q}_{k+1} \setminus \mathcal{Q}_{k} &  \subseteq  \{(A_{i_0}\cdots A_{i_{k}})^T\bar{Q}_i A_{i_0}\cdots A_{i_{k}}, \nonumber\\
& \quad \quad \bar{A}_{i_j}\in \bar{\mathcal{A}}, \forall i \in \mathcal{I}_p\} ,
\end{align}
and $|\mathcal{Q}_{k+1} \setminus \mathcal{Q}_{k}|  \le  M^{k+1}p$ for $k\in \mathbb{Z}^+_0$. Similarly, the update in (\ref{eqn:Hu0})-(\ref{eqn:Hlk}) is also adjusted as follows:
\begin{align}
\mathcal{H}^u_0 &= \{\bar{H}_i^u, i\in \mathcal{I}_m\}, \label{eqn:Husigma0}\\
\mathcal{H}^l_0 &= \{\bar{H}_i^l, i\in \mathcal{I}_m\}, \label{eqn:Hlsigma0}\\
\mathcal{H}^u_{k+1} &:= \{\bar{A}^T\bar{Q} \bar{A}: \bar{Q}\in \mathcal{H}^u_{k},\bar{A} \in \bar{\mathcal{A}}\}, k\in \Z_0^+ \label{eqn:Husigmak}\\
\mathcal{H}^l_{k+1} &:= \mathcal{H}^l_{k} \bigcup \{\bar{A} ^T\bar{Q} \bar{A}: \bar{Q}\in \mathcal{H}^l_{k},\bar{A} \in \bar{\mathcal{A}}\}, \label{eqn:Hlsigmak}
\end{align}
where  $\bar{H}_i^u$ and $\bar{H}_i^l$ are given in (\ref{eqn:barHui}) and (\ref{eqn:barHli}) respectively,  $|\mathcal{H}^u_{k}|  \le  M^{k}m$ and $|\mathcal{H}^l_{k}|  \le  \sum_{\ell=0}^k M^{\ell}m$ for $k\in \Z_0^+$.  With the assumption that $\rho(\mathcal{A})<1$, $A_{i_0}\cdots A_{i_{k}} \rightarrow 0$ as $k$ increases, for any $i_j \in\mathcal{I}_M, j=0,1,\cdots, k$. Following the arguments in Section \ref{sec:quad} \& \ref{sec:quadnl}, this implies that, there exists $k^*\in \mathbb{Z}_0^+$ such that, for any $Q\in (\mathcal{Q}_{k^*+1}\setminus \mathcal{Q}_{k^*}) \cup \mathcal{H}^u_{k^*+1}$, there exists  $\bar{Q}\in cone(\mathcal{Q}_{k^*} \cup \mathcal{H}^l_{k^*})$  that satisfies $Q\preceq \bar{Q}$. This implies that finite determination also holds for switched linear systems.

Due to the multiple modes in a switched system, at each iteration $k\in \Z^+_0$, we need to solve  at most   $M^{k+1}p$ LMI problems with  at most  $\sum_{\ell=0}^k M^{\ell}p+1$ variables at Algorithm \ref{algo:quad} and  at most  $M^{k+1}(p+m)$ LMI problems with  at most  $\sum_{\ell=0}^k M^{\ell}(p+m)+1$ variables at Algorithm \ref{algo:nl}. In this case, as $k$ increases, it may becomes necessary to remove redundancy using the formulation in (\ref{eqn:Okred}).

\subsection{Nonlinear systems with linear equivalents}
The proposed approach can be also extended to other special nonlinear systems. Consider the following nonlinear system
\begin{align}\label{eqn:fx}
x(t+1) = f(x(t)), \forall t \in \mathbb{Z}^+_0
\end{align}
where $x(t) \in \mathbb{R}^n$ and $f: \mathbb{R}^n \rightarrow \mathbb{R}^n$  is continuous with $f(0)=0$. The state is subject to
\begin{align}\label{eqn:Xnl}
x(t) \in X :=\{x: H_{i}(x) \le 1, i \in \mathcal{I}_m\}, \forall t \in \mathbb{Z}^+_0.
\end{align}
In the case of nonlinear systems, quadratic constraints are also included in (\ref{eqn:Xnl}). Similar to the linear case, the following assumptions are made.

\begin{assumption}\label{ass:stablenl}
System (\ref{eqn:fx}) is asymptotically stable at the origin in $X$, i.e., it converges to the origin for any initial state in $X$, and $f: \mathbb{R}^n \rightarrow \mathbb{R}^n$  is continuous with $f(0)=0$.
\end{assumption}

\begin{assumption}\label{ass:interiornl}
For all $i \in \mathcal{I}_m$, $H_{i}: \mathbb{R}^n \rightarrow \mathbb{R}$ is a continuous function with $H_i(0)=0$.  In addition, $X$ is compact.
\end{assumption}

The maximal \emph{CA-invariant} set of nonlinear systems can be defined in a similar way as shown in Section \ref{sec:pre}, although the computation is more complicated and difficult. Let the maximal \emph{CA-invariant} set of system (\ref{eqn:fx}) be denoted by $O_{\infty}^{nl}$, the same iterates can be used to compute $O_{\infty}^{nl}$
\begin{align}
O^{nl}_0 &:= X \label{eqn:O0nl}\\
O^{nl}_{k+1} &:= O^{nl}_k \bigcap \{x \in \mathbb{R}^n: f(x)\in O^{nl}_k\}, k \in \mathbb{Z}^+_0 . \label{eqn:Oknl}
\end{align}
Similar to the linear case, the maximal \emph{CA-invariant} set can be expressed as
\begin{align}
O_{\infty}^{nl}  : = \bigcap\limits_{k\in \mathbb{Z}^+_0} O^{nl}_{k}  =\{x: f^k(x)\in X,k\in \mathbb{Z}_0^+\}
\end{align}
where $f^k(x) = \underbrace{f\circ \cdots \circ f}_{k ~ times}(x)$ and $f^0(x) = x$.  With Assumptions \ref{ass:stablenl} and \ref{ass:interiornl}, the existence of $O_{\infty}^{nl}$ can be guaranteed and the algorithm above terminates in a finite time following similar arguments in Theorem 4.1 in \citep{ART:GT91} for the linear case. We believe such a result is already known or can be easily derived from some textbooks, see, e.g., \citep{BOO:A09}. However, we cannot find the exact reference in the literature. For completeness, we give the proof below.

\begin{proposition}\label{prop:Oinfnl}
 Consider System (\ref{eqn:fx}) with the constraint set $X$ as defined in (\ref{eqn:Xnl}), let $O^{nl}_k$ be defined in (\ref{eqn:O0nl})-(\ref{eqn:Oknl}) for any $k\in \mathbb{Z}^+$. With Assumptions \ref{ass:stablenl} and \ref{ass:interiornl}, the following properties hold: (i) For any $k\in \mathbb{Z}^+_0$, $O^{nl}_k$ is compact and contains the origin in its interior. (ii) There exists a finite $k^*$ such that
$
O_{k}^{nl}=O_{k^*}^{nl}
$
for all $k \ge k^*$ and $O_{\infty}^{nl} = O_{k^*}^{nl}$. 
\end{proposition}
Proof of Proposition \ref{prop:Oinfnl}: The proof is adapted from the proof of Theorem 4.1 in \citep{ART:GT91}. (i) From Assumption \ref{ass:interiornl}, we have $H_i(0)=0< 1$ for all $i\in \mathcal{I}_m$. Thus, from the definition of $X$ in (\ref{eqn:Xnl}), the origin is in the interior of $X$. For any $k\in \mathbb{Z}_0^+$, from the definition in (\ref{eqn:O0nl})-(\ref{eqn:Oknl}), $O^{nl}_k$ can be expressed as
$
O^{nl}_k = \{x\in \mathbb{R}^n:f^{\ell}(x) \in X,\ell \in \mathcal{I}_k \cup \{0\} \} = \{x\in \mathbb{R}^n: H_i(f^{\ell}(x))\le 1, i\in \mathcal{I}_m, \ell \in \mathcal{I}_k \cup \{0\} \}.
$
Under Assumption \ref{ass:stablenl}, we have that $f(x)$ is continuous with $f(0)=0$, which, together with the continuity of the functions $H_i(x)$,  implies that $H_i(f^{\ell}(x))$ is continuous with $H_i(f^{\ell}(0))=0<1$ for any $i\in \mathcal{I}_m$ and  $\ell \in \mathcal{I}_k \cup \{0\}$. This implies that the origin is in the interior of $O^{nl}_k$ for any $k\in \mathbb{Z}_0^+$. The compactness of $X$ and the continuity of $f(x)$ and $H_i(x)$ also imply that $O^{nl}_k$ is closed and bounded for any finite $k\in \mathbb{Z}^+$. According to the Heine–Borel theorem, they are also compact. (ii) Now, we show that $ O_{\infty}^{nl}=O^{nl}_{k^*}$ for some finite $k^*$. From Assumption \ref{ass:stablenl}, there exists a $k^*$ such that $f^{k^*}(x)\in X$ for any $x\in X$. We claim that $O^{nl}_{k^*}$ is an invariant set of System (\ref{eqn:fx}). We have to show that for any $x'\in O^{nl}_{k^*}$, $f(x')\in O^{nl}_{k^*}$. From the expression of $O^{nl}_{k^*}$ in (i), we can see that $x\in O^{nl}_{k^*}$ implies $f^{k}(x)\in X$ for all $k\in \mathcal{I}_{k^*}\cup \{0\}$. As the system is time-invariant, we know that $f^{k}(f(x'))\in X$ for $k\in \mathcal{I}_{k^*-1}\cup \{0\}$. From the fact that $f^{k^*}(x)\in X$ for any $x\in X$, we can see that $f^{k^*}(f(x'))\in X$, which implies that $f(x')\in O^{nl}_{k^*}$. This means that $O^{nl}_{k^*}$ is an invariant set. As defined in (\ref{eqn:Oknl}), from the invariance of $O^{nl}_{k^*}$, we get that $O^{nl}_{k^*+1} = O^{nl}_{k^*}$. Thus, it holds that
$
O_{k}^{nl}=O_{k^*}^{nl}
$
for all $k \ge k^*$, which implies that $O^{nl}_{k^*}=O^{nl}_{\infty}$. This completes the proof. $\Box$

Even though the existence of $O_{\infty}^{nl}$ is guaranteed, computing the exact $O_{\infty}^{nl}$ can be very challenging for general nonlinear systems, even when the nonlinear constraints satisfy Assumption \ref{ass:smooth}.  For this reason, we only consider a class of nonlinear systems that can be linearized by state transformation, see, e.g., \citep{ART:S82,ART:LM86,ART:MT12,INP:JT19}, for conditions for linearizability. While the state transformations in these papers are not necessarily diffeomorphisms, we make the following assumption for ease of discussion. 

\begin{assumption}\label{ass:diff}
There exists a diffeomorphism $T: \mathbb{R}^n \rightarrow \mathbb{R}^n$ such that System (\ref{eqn:fx}) can be transformed into a linear system \begin{align}\label{eqn:Ay}
y(t+1) = Ay(t), \forall t \in \mathbb{Z}^+_0
\end{align}
for some $A\in \mathbb{R}^{n\times n}$, $y(t)=T(x(t))$, with $T(0)=0$ and $f(x(t)) = T^{-1}(AT(x(t)))$.
\end{assumption}

An example of nonlinear systems that satisfy Assumption \ref{ass:diff} will be given in the next section. The linearized system (\ref{eqn:Ay}) is subject to the following constraints
\begin{align}
y(t) \in Y := T(X), \forall t \in \mathbb{Z}^+_0
\end{align}
with
$
T(X) =\{y\in \mathbb{R}^n: H_{i}(T^{-1}(y)) \le 0, i \in \mathcal{I}_m\}.
$
With the state transformation, it is possible to compute the maximal \emph{CA-invariant} of System (\ref{eqn:fx}) by computing the maximal \emph{CA-invariant} set of the linearized system (\ref{eqn:Ay}). Let $O_{\infty}^{Y}$ denote the maximal \emph{CA-invariant} set of the linearized system (\ref{eqn:Ay}). Suppose $Y$ satisfies Assumption \ref{ass:smooth}, $O_{\infty}^{Y}$ can be computed using Algorithm \ref{algo:nl}. The equivalence between the invariant sets of System (\ref{eqn:fx}) and  System (\ref{eqn:Ay}) can be easily established. In many real applications, we will need to deal with systems with nonlinear dynamics and linear (or box) constraints. In this case, $Y$ will often satisfy Assumption \ref{ass:smooth} (when $T^{-1}(y)$ is continuously differentiable with Lipschitz gradient), although it is not guaranteed.

\begin{remark}
From the discussion above, we can see that it is possible to compute the maximal \emph{CA-invariant} set of nonlinear systems using their linear equivalents in some cases. However, the problem of computing linear equivalents for nonlinear systems is nontrivial and it is out of the scope of this paper. For a detailed discussion, we refer readers to a recent paper \citep{INP:WJO20} and the references therein.
\end{remark}


\section{Illustrative examples}\label{sec:num}
\begin{example}\label{exam:Lin1}
We consider the linear system studied in \citep[Example 1]{INP:AJ16} with $A=[1.0216 ~ 0.3234;-0.6597 ~ 0.5226]$. The constraint set is  the unit circle given by  $\Omega_1:=\{x\in \mathbb{R}^2:x^Tx  \le 1\}$ and $\Theta=\mathbb{R}^n$. Algorithm \ref{algo:quad} is used to obtain the maximal \emph{CA-invariant} set and the result is given in Figure \ref{fig:ex1}. It can been seen from Figure \ref{fig:ex1} that Algorithm \ref{algo:quad}  terminates at $t^*=3$. For the same setting, the algorithm in \citep{INP:AJ16} takes $6$ iterations.

\begin{figure}[h]
\centering
\includegraphics[width=0.5\linewidth]{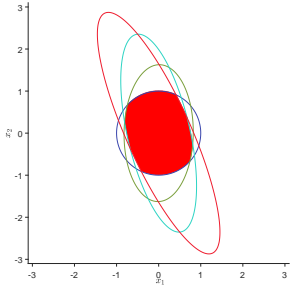}
\caption{The maximal \emph{CA-invariant} set $O_{\infty}(O_{6})$ of Example \ref{exam:Lin1} with $\Omega = \Omega_1$ and $\Theta=\mathbb{R}^n$.}
\label{fig:ex1}
\end{figure}

We consider the same dynamical system in Example \ref{exam:Lin1} with additional quadratic constraints. Let the quadratic constraint set be $\Omega_2:=\{x\in \mathbb{R}^2:x^Tx  \le 1, 2x_1^2-x_2^2+0.4x_1x_2 \le 1, (x_1+0.5)^2+x_2^2 \ge \frac{1}{16},(x_1-0.5)^2+x_2^2 \ge \frac{1}{16}\}$. Note that there are $4$ quadratic constraints and that this set is nonconvex. Again, we use Algorithm \ref{algo:quad} to compute the maximal \emph{CA-invariant} set and it terminates at $t^*=8$. The set is shown in Figure \ref{fig:ex2}. Trajectories are also shown to verify set invariance of the disconnected regions.

\begin{figure}[h]
  \centering
  \begin{tabular}{cc}
  \subcaptionbox{ \label{fig:2a}}{\includegraphics[width=1.5in]{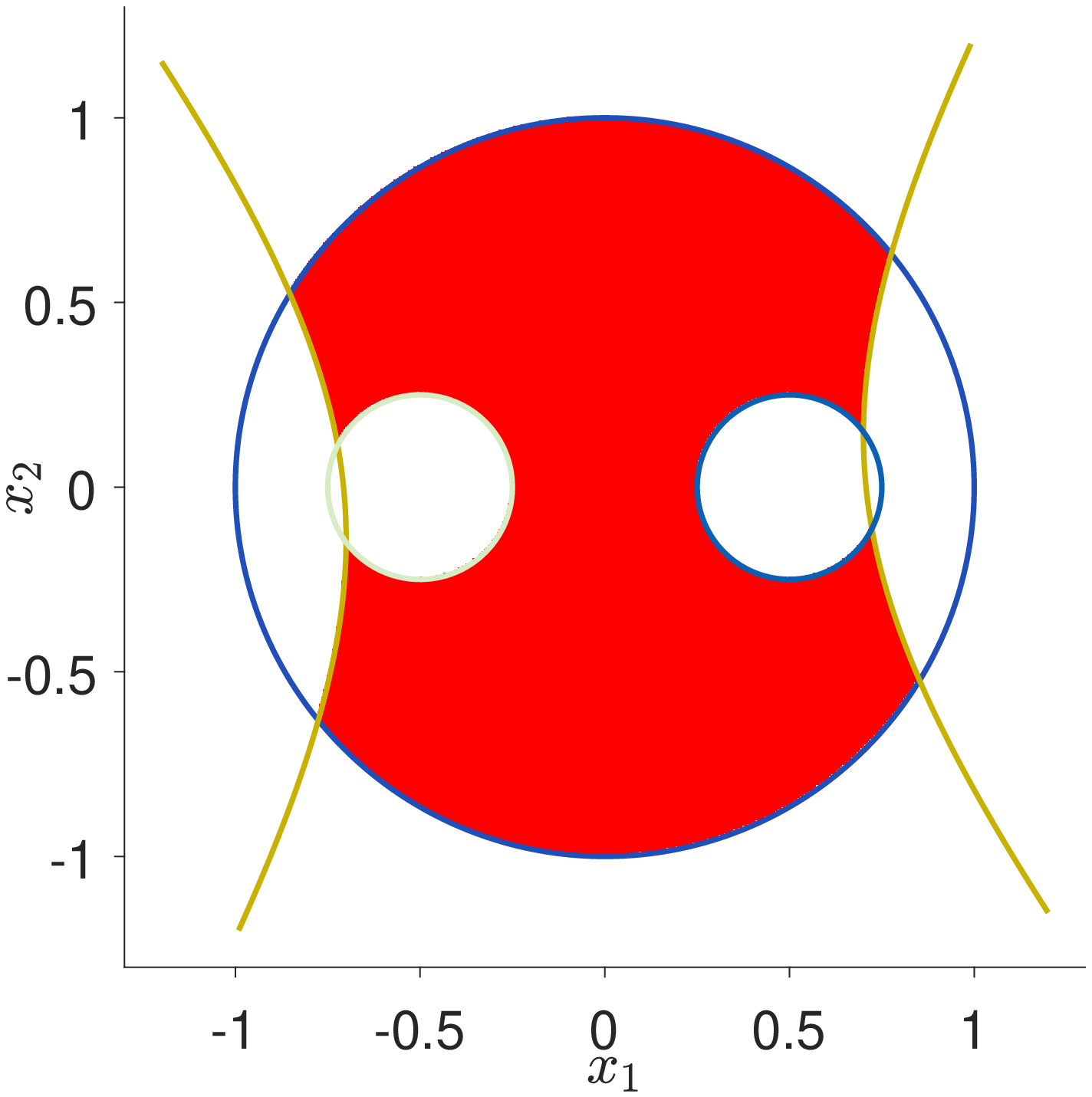}} & \subcaptionbox{ \label{fig:2b}}{\includegraphics[width=1.5in]{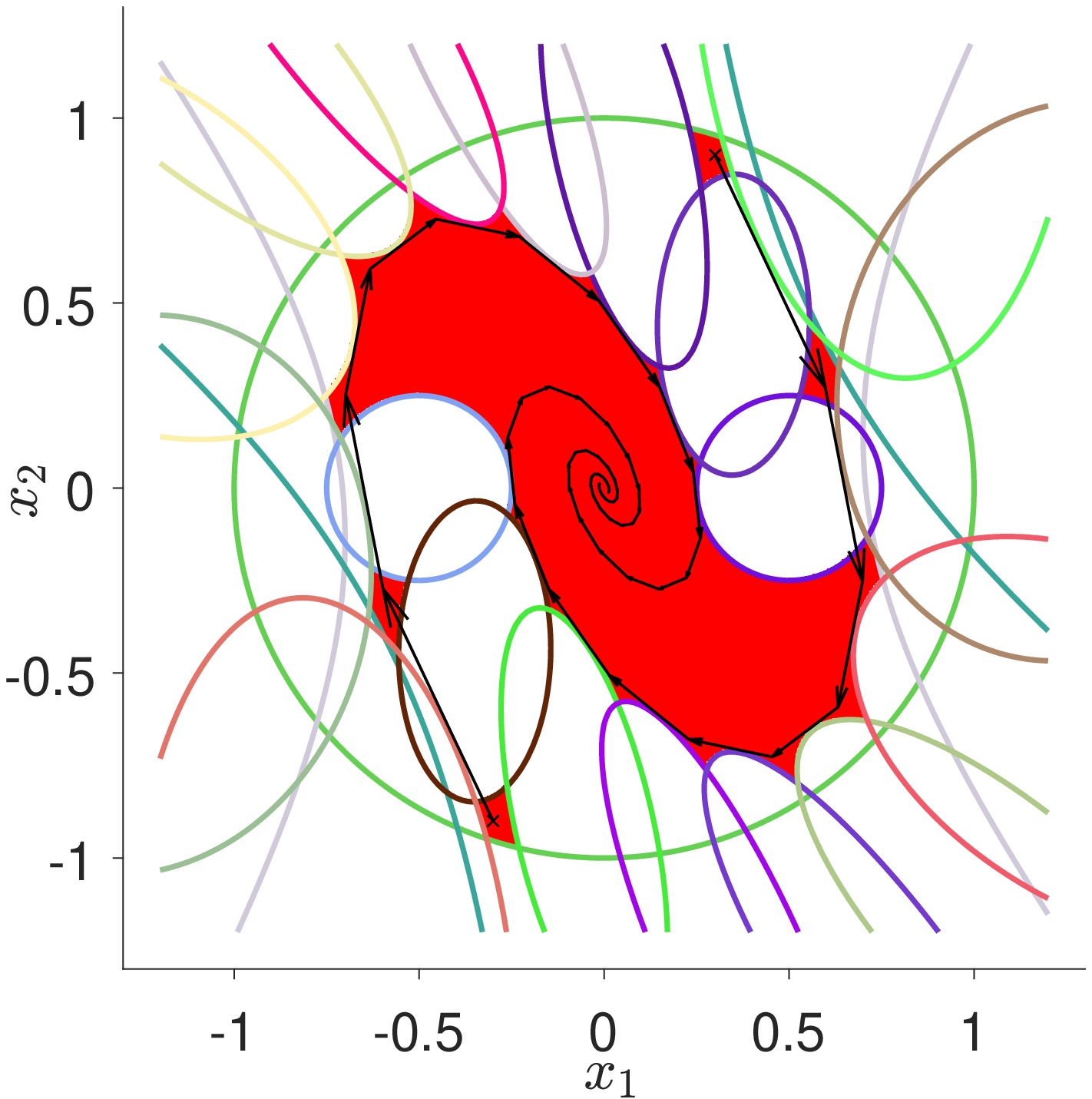}}\\
  \end{tabular}
\caption{The maximal \emph{CA-invariant} set of Example \ref{exam:Lin1} with $\Omega=\Omega_2$ and $\Theta=\mathbb{R}^n$: (a) shows the set $\Omega$, and (b) shows the maximal \emph{CA-invariant} set $O_{\infty}(O_{8})$.}
\label{fig:ex2}
\end{figure}

Additionaly, we also consider a nonlinear constraint, which is beyond the class of constraints that the approach in \citep{INP:AJ16} is able to handle. Let $\Theta=\Theta_1:=\{x\in \mathbb{R}^2: H_1(x):=\sqrt{x_1^2+x_2^2+1}+2x_1+2x_2-2 \le 0\}$. It is easy to verify that Assumption \ref{ass:smooth} is satisfied with $H_{1}^{\nabla}=[2 ~2]^T$ and $L_1=1$. Using Algorithm \ref{algo:nl}, the maximal \emph{CA-invariant} set can be obtained with $t^*=8$ as shown in Figure \ref{fig:ex3}.

\begin{figure}[h]
  \centering
  \begin{tabular}{cc}
  \subcaptionbox{ \label{fig:3a}}{\includegraphics[width=1.5in]{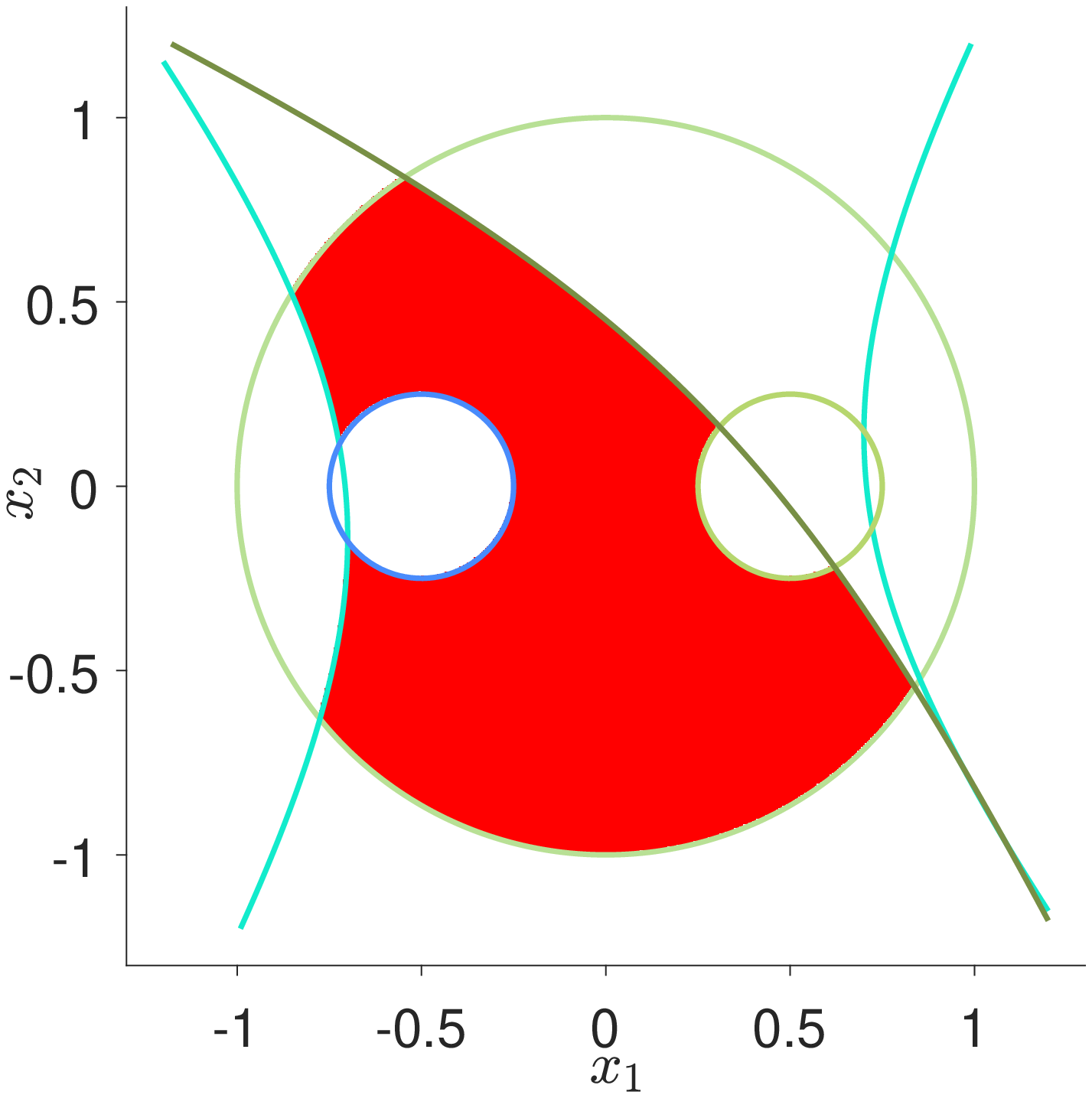}} & \subcaptionbox{ \label{fig:3b}}{\includegraphics[width=1.5in]{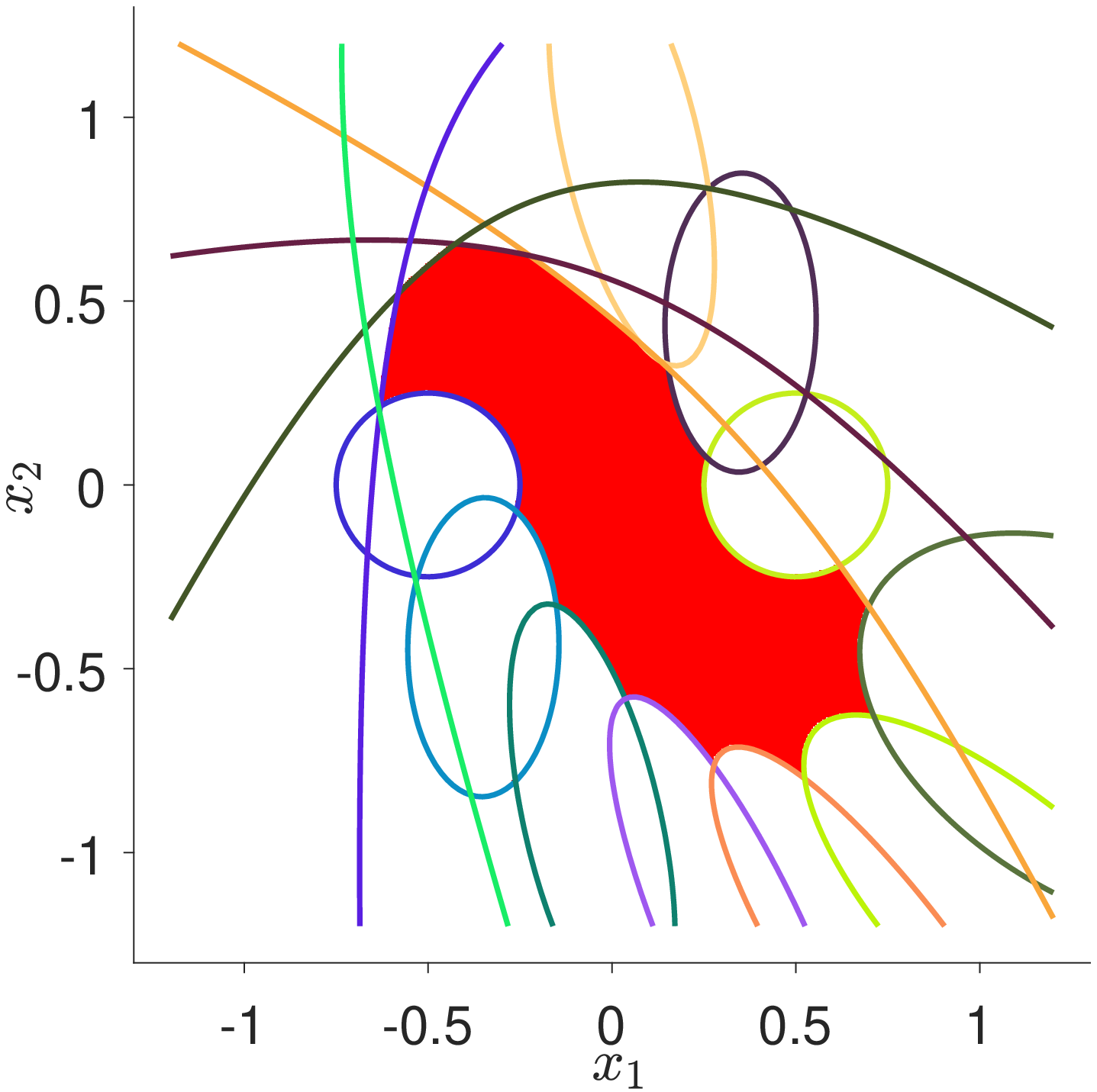}}\\
  \end{tabular}
\caption{The maximal \emph{CA-invariant} set of Example \ref{exam:Lin1} with $\Omega=\Omega_2$ and $\Theta=\Theta_1$: (a) shows the set $\Omega \bigcap \Theta$, and (b) shows the maximal \emph{CA-invariant} set $O_{\infty}(O_{8})$.}
\label{fig:ex3}
\end{figure}
\end{example}

\begin{example}\label{exam:wiener}
We consider an autonomous Wiener system, which consists of  a linear dynamical system and a nonlinear static system  (see \citep{ART:M19} for details on autonomous Wiener systems), as shown in Figure \ref{fig:wiener}, with $A = [0.5 ~ 0.7;-0.7 ~ 0.5]$, $C=[1 ~ -1]$ and $g(v) = v+v^2+v^3-v^4$. The constraints are given by: $\Omega=\{x\in \mathbb{R}^2: x_1^2+x_2^2 \le 2.5\}$ and $\Theta=\{x\in \mathbb{R}^2: -2\le g(Cx) \le 2\}$.

\tikzstyle{int}=[draw, fill=blue!20, minimum size=2em]
\begin{figure}[h]
  \centering
\begin{tikzpicture}[node distance=4.5cm,auto,>=latex']
    \node [int] (a) {$x(t+1) = Ax(t)$};
    \node [int] (c) [right of=a] {$g(v(t))$};
    \node [coordinate] (end) [right of=c, node distance=2cm]{};
    \path[->] (a) edge node {$v(t)=Cx(t)$} (c);
    \draw[->] (c) edge node {$y(t)$} (end) ;
\end{tikzpicture}
\caption{A discrete-time autonomous Wiener model}
\label{fig:wiener}
\end{figure}
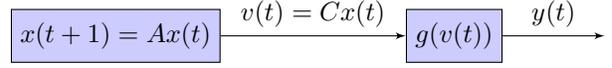

The output $g(Cx)$ can be rewritten as
\begin{align}
g(Cx) =\left( \begin{array}{c}
x_1\\
x_2\\
x_1x_2\\
x_1^2\\
x_2^2
\end{array} \right)^T P\left( \begin{array}{c}
x_1\\
x_2\\
x_1x_2\\
x_1^2\\
x_2^2
\end{array} \right) + 2F^T x
\end{align}
with 
$$
P = \left( \begin{array}{ccccc}
1 & -1 & -1.5 & 0.5 & 1.5\\
-1 &  1 &  0 &  0   & -0.5\\
-1.5 &  0 &   -6  &  2 &  2\\
0.5 &  0  &   2 &   -1 &   0\\
1.5 &  -0.5 &  2 &  0 &   -1\\
\end{array} \right) \textrm{ and }
$$
$F = [0.5 ~-0.5]^T$. The lifted system $\tilde{A}$ in (\ref{eqn:zAtilde}) becomes
$$
\tilde{A} = \left( \begin{array}{ccccc}
0.5 &  0.7 &   0   &      0  &   0\\
   -0.7 &    0.5 &   0  &  0  &  0\\
         0   &      0 &  -0.24 &   -0.35 &  0.35\\
         0   &     0 &    0.7 &    0.25 &    0.49\\
         0   &     0 &   -0.7 &    0.49 &    0.25\\
\end{array} \right).
$$
With the inequality $x_1^2+x_2^2\le 2.5$, it can be easily verified that 
$$
\left( \begin{array}{c}
x_1\\
x_2\\
x_1x_2\\
x_1^2\\
x_2^2
\end{array} \right)^T\left( \begin{array}{c}
x_1\\
x_2\\
x_1x_2\\
x_1^2\\
x_2^2
\end{array} \right) \le 8.75.
$$
Then, the constraint set for the lifted system is
$
X_z = \{z\in \mathbb{R}^5: z^T[I_2 ~ \pmb{0}]^T[I_2 ~ \pmb{0}]z \le 2.5, z^TPz+2F^T [I_2 ~ \pmb{0}]z \le 2,-z^TPz-2F^T [I_2 ~ \pmb{0}]z  \le 2, z^Tz \le 8.75\}.
$
Finally, the lifted maximal \emph{CA-invariant}  set $O_{\infty}^z$ can be obtained using Algorithm \ref{algo:quad}, which terminates at $k=5$. According to Proposition \ref{prop:lift}, the   maximal \emph{CA-invariant}  set of the original system can be given by $O_{\infty}=\{x\in \mathbb{R}^2: (x_1,x_2,x_1x_2,x_1^2,x_2^2) \in O_{\infty}^z \}$, which is shown in Figure \ref{fig:wienerOinf}. Again, a trajectory is given to verify set invariance of the disconnected regions.

\begin{figure}[h]
  \centering
  \begin{tabular}{cc}
  \subcaptionbox{ \label{fig:wienera}}{\includegraphics[width=1.5in]{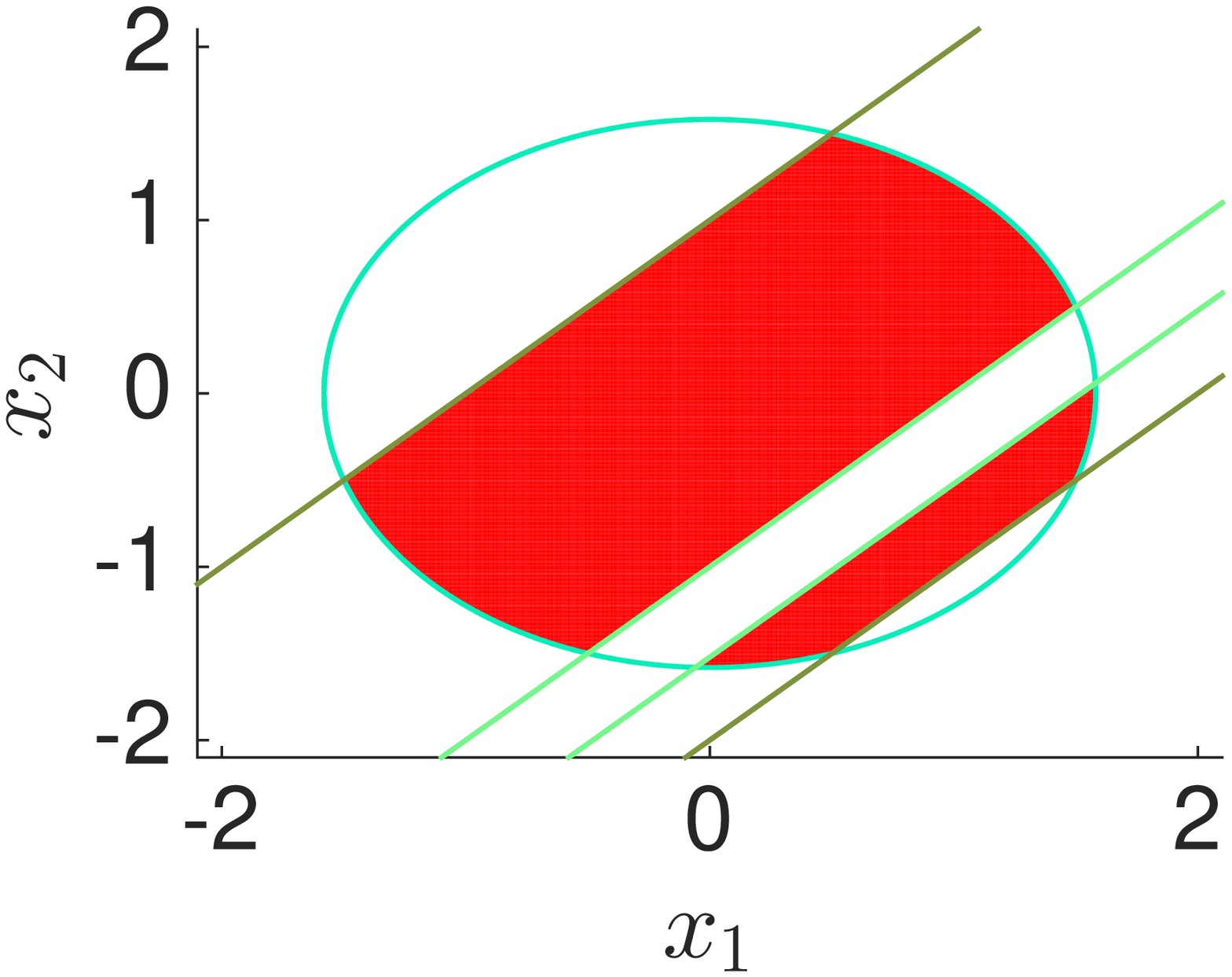}} & \subcaptionbox{ \label{fig:wienerb}}{\includegraphics[width=1.5in]{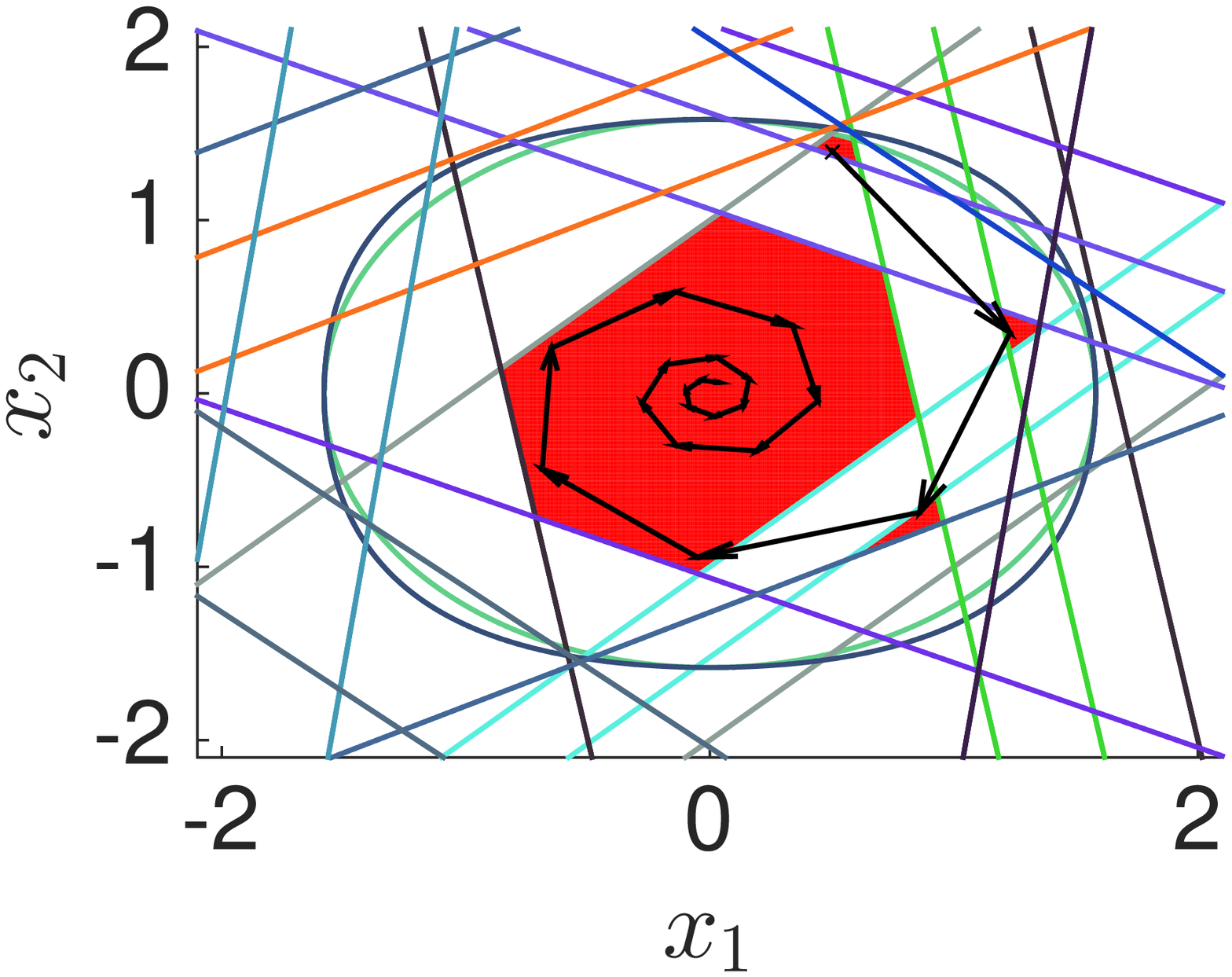}}\\
  \end{tabular}
\caption{The maximal \emph{CA-invariant} set $O_{\infty}$ of the Wiener system: (a) shows the set $\Omega \bigcap \Theta$, and (b) shows the set $O_{\infty}$.}
\label{fig:wienerOinf}
\end{figure}

\end{example}

\begin{example}\label{exam:rand}
Now, we evaluate the proposed approach on switched linear systems of different sizes. As we have already seen in Example \ref{exam:Lin1}, compared with the lifting approach in \citep{INP:AJ16}, our approach takes fewer iterations for the same setting. In this example, we will make more comparison experiments in more difficult situations. Consider a switched linear system (\ref{eqn:Asigma}) with $\mathcal{A} = \{A_1,A_2\}$, which are randomly generated. To make sure that $\rho(\mathcal{A})<1$ is satisfied, we first generate matrices $\hat{A}_1$ and $\hat{A}_2$ whose elements are sampled independently and identically from the uniform distribution between $-1$ and $1$. Then, we compute the JSR $\rho(\{\hat{A}_1,\hat{A}_2\})$ (or an upper bound) using the JSR toolbox \citep{INP:VHJ14}. Finally, we let
$$
A_1 = \frac{\hat{A}_1}{\rho(\{\hat{A}_1,\hat{A}_2\})+\epsilon}, ~~A_2 = \frac{\hat{A}_2}{\rho(\{\hat{A}_1,\hat{A}_2\})+\epsilon}, 
$$
where $\epsilon>0$. With this choice of $\{A_1,A_2\}$, the condition that $\rho(\mathcal{A})<1$ is satisfied for any $\epsilon>0$. In the simulation, we set $\epsilon=0.1$. The constraint set is given by $X=\{x\in \mathbb{R}^n: x^Tx \le 1, x^TQ_ax+2q_a^T x \le 1,x^TQ_bx+2q_b^Tx \le 1 \}$, where the symmetric matrices $Q_a,Q_b\in \mathbb{S}^n$ and the vectors $q_a,q_b \in \mathbb{R}^n$ are also randomly generated. We then use Algorithm \ref{algo:quad} with the modifications in (\ref{eqn:Osigma0})-(\ref{eqn:Osigmak}) and (\ref{eqn:mathcalQsigma0})-(\ref{eqn:mathcalQsigmak}) to compute $O_{\infty}$. Let $\mathcal{N}_{iter}$ denote the number of iterations and $\mathcal{N}_{const}$ denote the number of constraints in the expression of $O_{\infty}$ (or equivalently $O_{\mathcal{N}_{iter}}$) after removing redundancy by solving (\ref{eqn:Okred}).  Note that different approaches may result in different descriptions of $O_{\infty}$ in the presence of nonlinear constraints though the set $O_{\infty}$ is fixed, because identifying redundant nonlinear constraints requires us to solve non-convex problems, see Problems (\ref{eqn:gmax}) and (\ref{eqn:hmax}). The comparison with the lifting approach in \citep{INP:AJ16} is made in terms of the number of iterations and the number of constraints in the expression of $O_{\infty}$. Similarly, let $\mathcal{N}'_{iter}$ and $\mathcal{N}'_{const}$ denote the number of iterations and the number of constraints respectively in \citep{INP:AJ16}. The approach in \citep{INP:AJ16} lifts the system into a $\frac{(n+3)n}{2}$-dimensional system, where the quadratic constraints become linear constraints, while our approach does not have to lift the system as the constraints are quadratic.  For the lifted system of \citep{INP:AJ16}, all the sets from (\ref{eqn:Ok}) are polyhedra and we can remove redundancy by solving linear optimization problems according to the extended Farkas' lemma \citep{BOO:S86,ART:Bla99}. The computation of polyhedra is implemented with the Multi-Parametric Toolbox \citep{INP:MPT3}, which allows to remove redundancy efficiently.

We take $20$ realizations of the dynamics and the constraints and compute the mean values of $\mathcal{N}_{iter}$, $\mathcal{N}_{const}$, $\mathcal{N}'_{iter}$ and $\mathcal{N}'_{const}$, denoted by $\overline{\mathcal{N}_{iter}}$, $\overline{\mathcal{N}_{const}}$, $\overline{\mathcal{N}'_{iter}}$, and $\overline{\mathcal{N}'_{const}}$ respectively. The results are shown in Table \ref{table:Oinfcom}. When $n>5$, the approach in \citep{INP:AJ16} is not conducted as it takes too much time. As we can see in Table \ref{table:Oinfcom}, the proposed approach converges faster and produces a tighter expression of $O_{\infty}$ with a smaller number of constraints.

\begin{table}[h]
\renewcommand{\arraystretch}{1.4}
\centering
\begin{tabular}{|c|c|c|c|c|}
\hline
$n$ & $\overline{\mathcal{N}_{iter}}$  & $\overline{\mathcal{N}_{const}}$  & $\overline{\mathcal{N}'_{iter}}$  & $\overline{\mathcal{N}'_{const}}$  \\ \hline
$2$ & $2.25$  & $6.65$  & $6.45$  & $30.65$  \\ \hline
$3$ & $3.85$  & $15.1$ & $7.9$  & $70.6$ \\ \hline
$4$ & $5.55$  & $29.5$ & $11.45$  & $208.15$ \\ \hline
$5$ & $6.55$  & $40.2$ & $12.55$  & $328.25$ \\ \hline
$6$ & $7.85$  & $64.65$ & -  & - \\ \hline
$10$ & $9.6$  & $130.65$ & -  & - \\ \hline
$20$ & $13.4$  & $467.15$ & -  & - \\ \hline
$30$ & $14.35$  & $1.23\times 10^3$ & -  & - \\ \hline
\end{tabular}
\caption{Comparison with the lifting approach \citep{INP:AJ16} for Example \ref{exam:rand} of different sizes with $20$ realizations.}
\label{table:Oinfcom}
\end{table}

\end{example}

\begin{example}\label{exam:nonl}
In the rest of this section, we consider the following nonlinear system
\begin{align}
\begin{aligned}
x_1(t+1) &= 2(x_1(t))^2+x_2(t),\\
x_2(t+1) &= -2\left(2(x_1(t))^2+x_2(t)\right)^2-0.8x_1(t).
\end{aligned}
\end{align}
The state constraint set is given by $X:=\{x\in \R^2: |x_1|\le 1, |x_2|\le 1\}$. There exists a diffeomorphism $y=T(x)$,
\begin{align}
T(x) = \left( \begin{array}{c}
x_1\\
2x_1^2+x_2
\end{array}\right),
\end{align}
such that the nonlinear system can be linearized into
\begin{align}
y(t+1) = \left(\begin{array}{cc}
0 & 1\\
-0.8 & 0
\end{array}\right)y(t).
\end{align}
With the state transformation $T(x)$, the state constraint set of the linearized system can be given by $Y:=\{y\in \R^2: |y_1|\le 1, y_2-2y_1^2\le 1, 2y_1^2-y_2\le 1\}$. As a result, we get a linear system with quadratic constraints and the constraint set $Y$ is bounded. Using Algorithm \ref{algo:quad}, the maximal \emph{CA-invariant} set of the linearized system can be computed and it takes $3$ iterations. The set is shown in Figure \ref{fig:Lin}.

\begin{figure}[h]
  \centering
  \begin{tabular}{cc}
  \subcaptionbox{ \label{fig:Lina}}{\includegraphics[width=1.2in]{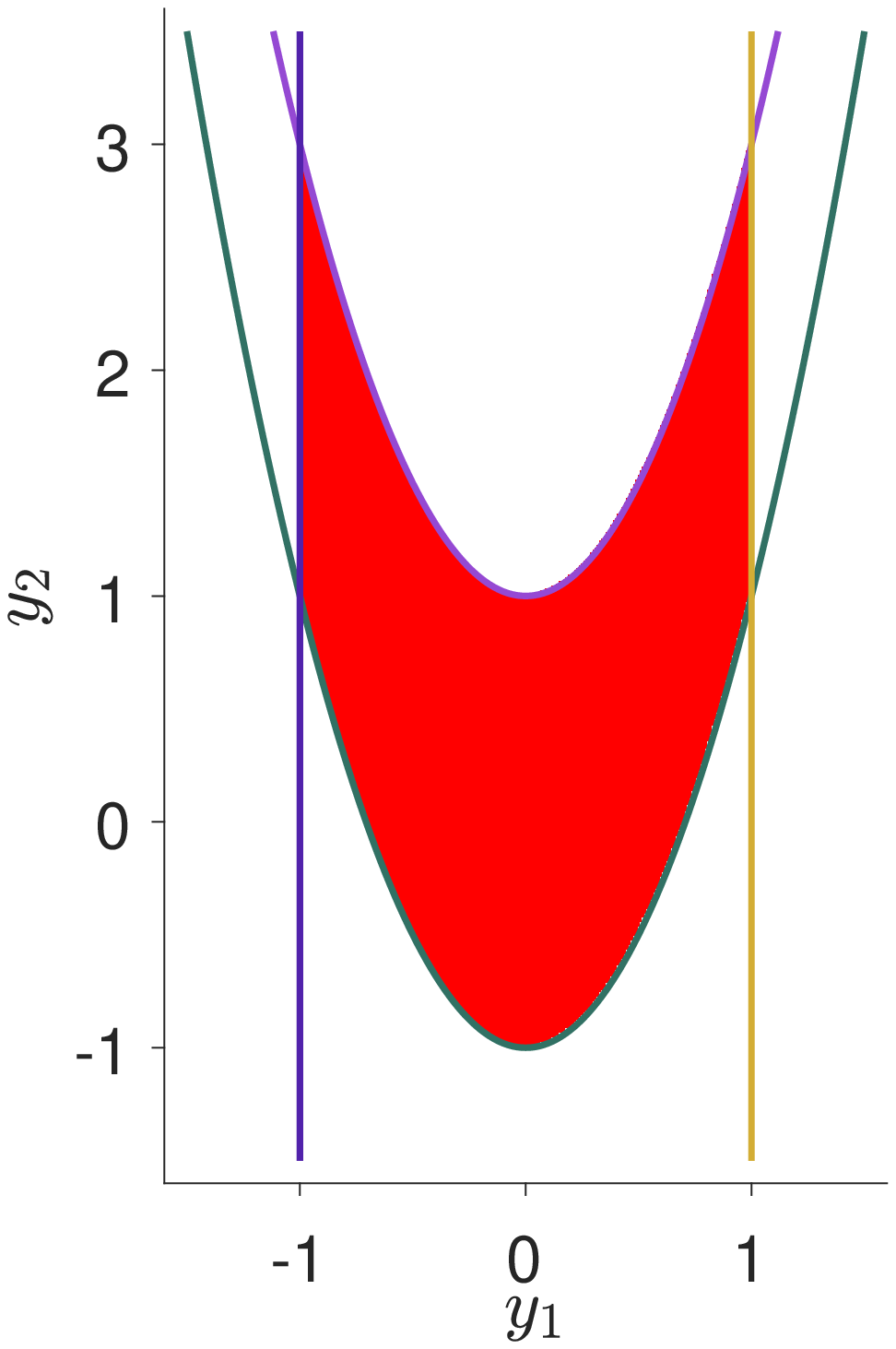}} & \subcaptionbox{ \label{fig:Linb}}{\includegraphics[width=1.2in]{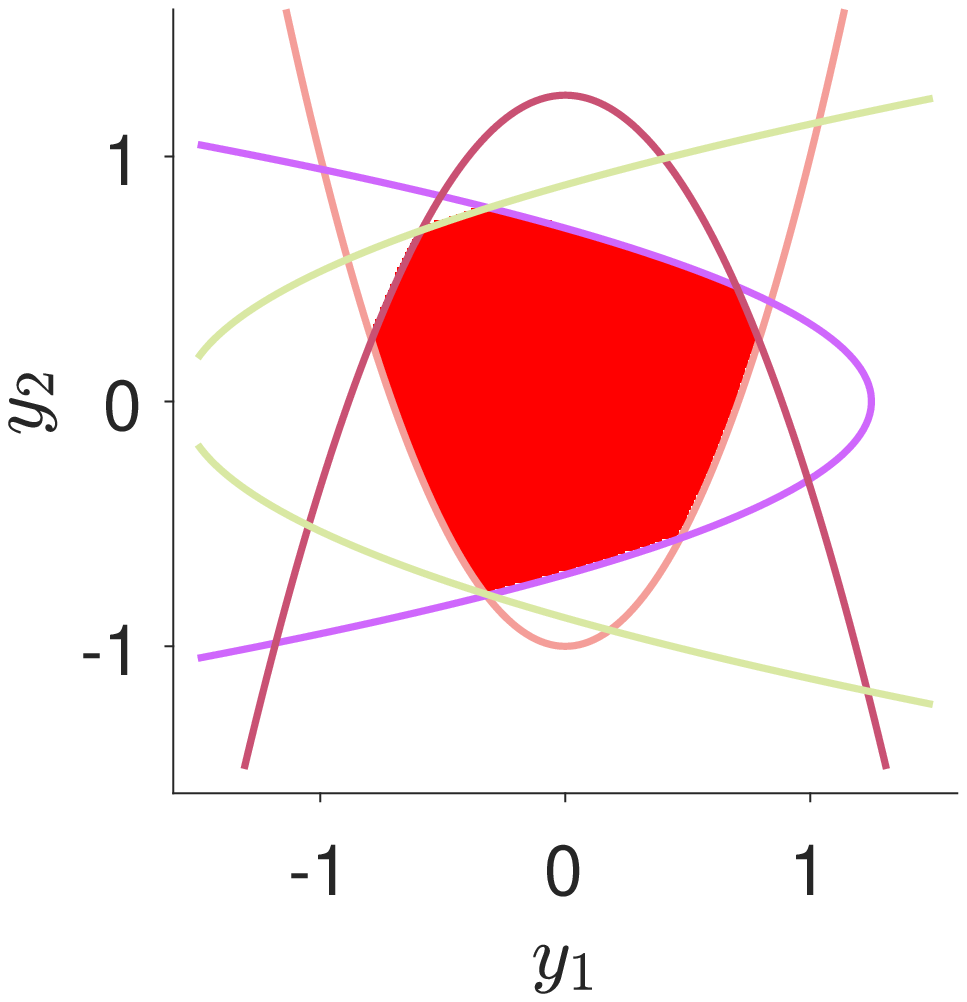}}\\
  \end{tabular}
\caption{The maximal \emph{CA-invariant} set of the linearized system of Example \ref{exam:nonl}: (a) shows the set $Y$ and (b) shows the maximal \emph{CA-invariant} set $O_{\infty}^{Y}$.}
\label{fig:Lin}
\end{figure}

Using the inverse mapping $x=T^{-1}(y)$,
\begin{align}
T^{-1}(y) = \left( \begin{array}{c}
y_1\\
y_2 - 2y_1^2
\end{array}\right),
\end{align}
the maximal \emph{CA-invariant} set of the original nonlinear system can be obtained and is shown in Figure \ref{fig:NL}.

\begin{figure}[h]
  \centering
  \begin{tabular}{cc}
  \subcaptionbox{ \label{fig:NLa}}{\includegraphics[width=1.5in]{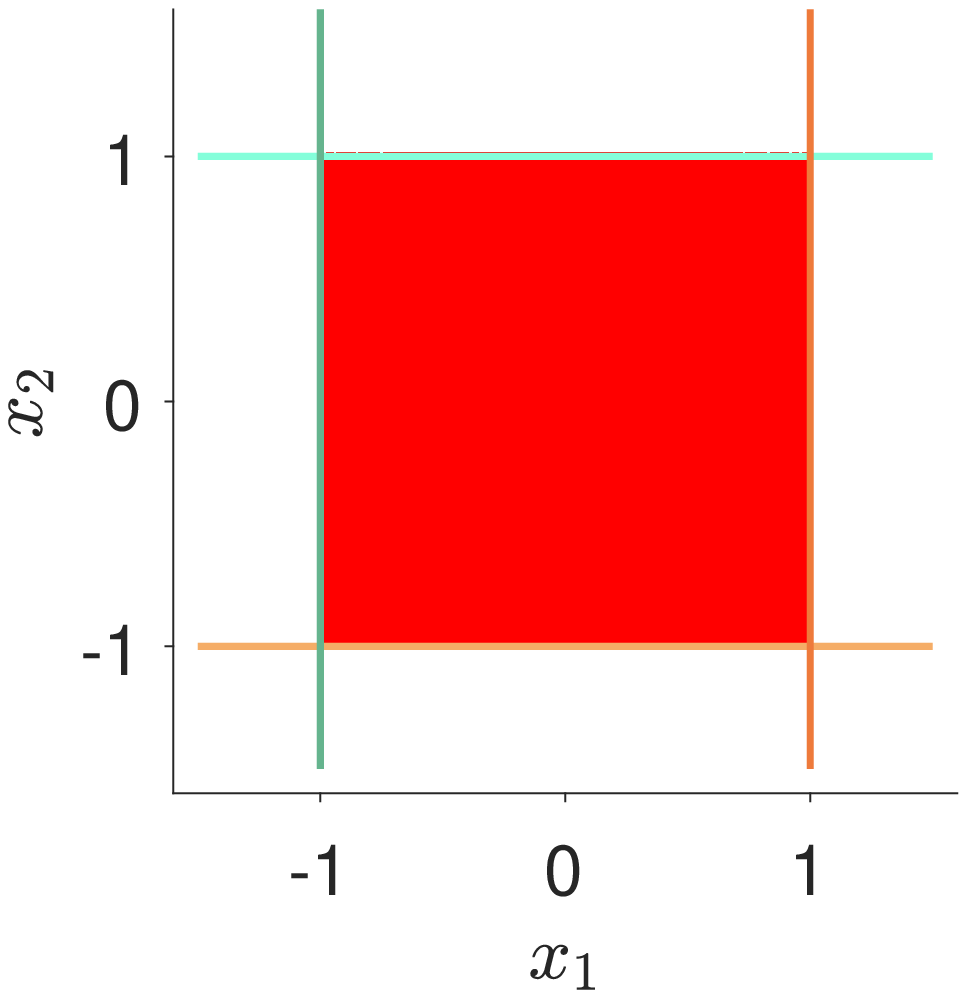}} & \subcaptionbox{ \label{fig:NLb}}{\includegraphics[width=1.5in]{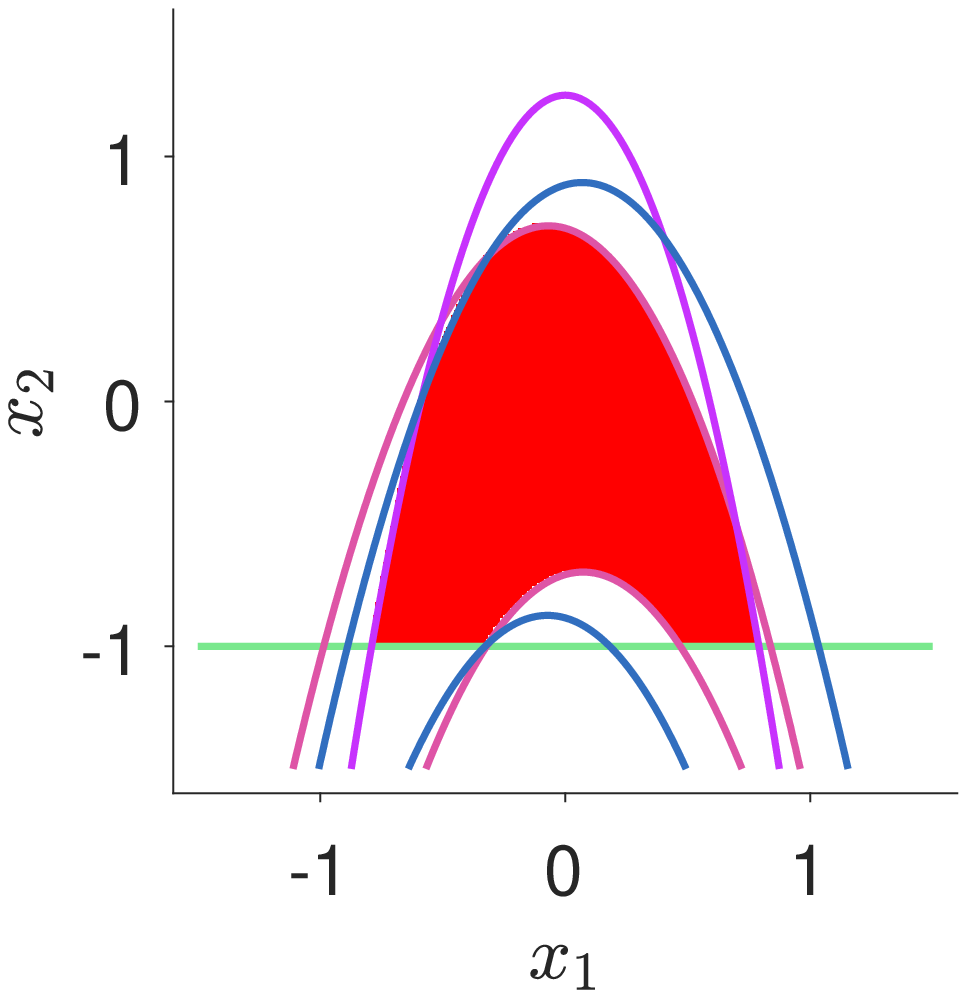}}\\
  \end{tabular}
\caption{The maximal \emph{CA-invariant} set of Example \ref{exam:nonl}: (a) shows the set $X$ and (b) shows the maximal \emph{CA-invariant} set $O_{\infty}^{nl}$.}
\label{fig:NL}
\end{figure}

\end{example}
\section{Conclusions}\label{sec:con}
We have studied the computation of the maximal \emph{CA-invariant} set of discrete-time linear systems subject to a class of non-convex constraints that admit quadratic lower and upper bounds.  By the use of these quadratic bounds, we have derived a sufficient condition for set invariance, which can be expressed as a set of LMIs. Based on this sufficient condition, a new algorithm is presented by solving a number of convex problems with only one LMI constraint at every iteration. Under mild assumptions, finite convergence to the exact maximal \emph{CA-invariant} set can be guaranteed.  This algorithm can be extended to switched linear systems and some special nonlinear systems that admit linear equivalents.  To illustrate the performance of the proposed algorithm, we have presented several numerical examples and made comparison with an existing approach, which is capable of computing the exact maximal \emph{CA-invariant} set of switched linear systems subject to semi-algebraic constraints. For the same setting, we show that our approach converges faster with a tighter expression of the maximal \emph{CA-invariant} set.

\appendix
\section*{Appendix}
\subsection*{Proof of Proposition  \ref{lem:linear}}
In the case of linear constraints, $O_k$ is a polyhedral set for any $k\in \mathbb{Z}_0^+$. It is clear from Lemma \ref{lem:sprocedure} that $\mathcal{R}_{\max}(k)\le 0$ implies $O_{k+1} = O_k$. We only need to show $O_{k+1} = O_k$ implies $\mathcal{R}_{\max}(k)\le 0$. From (\ref{eqn:OkA}), $O_{k+1} = O_k$ if and only if $O_k \subseteq  \{x\in \mathbb{R}^n: A^{k+1}x\in X\}$. From the extended Farkas' lemma \citep{BOO:S86,ART:Bla99}, for any $k\in \mathbb{Z}_0^+$,  $O_k \subseteq  \{x\in \mathbb{R}^n: A^{k+1}x\in X\}$ if and only if there exists a non-negative matrix $S\in \mathbb{R}^{p\times (k+1)p}$ such that,
\begin{align}
S\left(\begin{array}{c}
q^T\\
q^TA\\
\vdots\\
q^TA^k
\end{array}\right) &= q^TA^{k+1}, \label{eqn:PGA}\\
\sum\limits_{j=1}^{p(k+1)}S_{(i,j)} &\le 1, \forall i \in \mathcal{I}_p. \label{eqn:P1}
\end{align}
Suppose there exists a non-negative matrix $S\in \mathbb{R}^{p\times (k+1)p}$ satisfying (\ref{eqn:PGA}) and (\ref{eqn:P1}) for some $k\in \mathbb{Z}^+_0$, by simple manipulations, we can see that  $q_i^TA^{k+1} =\sum_{\ell=0}^{k}  \sum_{j=1}^{p}  S(i,p\ell+j) q_j^TA^\ell$ and $\sum_{\ell=0}^{k} \sum_{j=1}^{p} S(i,p\ell+j)\le 1$ for any $i\in \mathcal{I}_p$, which implies that
\begin{align}
&\left(\begin{array}{cc} 0 & (A^{k+1})^Tq_i\\ q_i^TA^{k+1} & -1 \end{array}\right) \nonumber\\
&- \sum_{\ell=0}^{k} \sum_{j=1}^{p} S(i,p\ell+j)  \left(\begin{array}{cc} 0 & (A^{\ell})^Tq_i\\ q_i^TA^{\ell} & -1 \end{array}\right) \nonumber\\
=&\left(\begin{array}{cc} 0 & 0\\ 0 & \sum_{\ell=0}^{k} \sum_{j=1}^{p} S(i,p\ell+j) -1 \end{array}\right) \preceq 0.
\end{align}
This means that $\mathcal{R}(Q,\mathcal{Q}_k)\le 0$ for any $Q \in \mathcal{Q}_{k+1} \setminus \mathcal{Q}_k$. Hence, $\mathcal{R}_{\max}(k)\le 0$. $\Box$

\bibliographystyle{dcu}
\bibliography{Reference}

\end{document}